\documentclass[12pt]{article}

\usepackage{setspace,graphicx,epstopdf,amsmath,amsfonts,amssymb,amsthm,versionPO}
\usepackage{marginnote,datetime,enumitem,rotating,fancyvrb}
\usepackage{hyperref,float}
\usepackage{booktabs}
\usepackage[T1]{fontenc}
\usepackage{natbib}
\usepackage{siunitx}
\usepackage{threeparttable}
\usepackage{subcaption}
\usepackage{caption}
\usepackage{dcolumn}
\usepackage{tabularx}
\usepackage{multirow}
\usepackage{adjustbox}
\usepackage{pdflscape}
\usepackage{longtable}
\usepackage{tabu}
\usepackage{lscape}
\usepackage{afterpage}
\usepackage{csquotes}

\newcolumntype{d}{D{.}{.}{4}}
\usdate

\excludeversion{notes}        % Include notes?
\includeversion{links}          % Turn hyperlinks on?

\iflinks{}{\hypersetup{draft=true}}

\ifnotes{%
\usepackage[margin=1in,paperwidth=10in,right=2.5in]{geometry}%
\usepackage[textwidth=1.4in,shadow,colorinlistoftodos]{todonotes}%
}{%
\usepackage[margin=1in]{geometry}%
\usepackage[disable]{todonotes}%
}

\makeatletter\let\chapter\@undefined\makeatother % Undefine \chapter for todonotes

\newcommand*{\SuperScriptSameStyle}[1]{%
  \ensuremath{%
    \mathchoice
      {{}^{\displaystyle #1}}%
      {{}^{\textstyle #1}}%
      {{}^{\scriptstyle #1}}%
      {{}^{\scriptscriptstyle #1}}%
  }%
}

\newcommand*{\oneS}{\SuperScriptSameStyle{*}}
\newcommand*{\twoS}{\SuperScriptSameStyle{**}}
\newcommand*{\threeS}{\SuperScriptSameStyle{*{*}*}}

\newcommand{\qlet}{\raisebox{-1pt}{\includegraphics[scale=0.042]{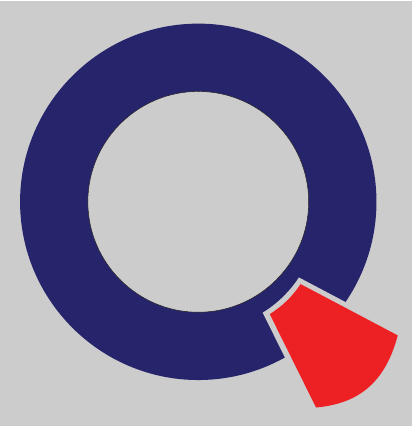}}}
\usepackage{indentfirst} % Indent first sentence of a new section.
\usepackage{jfe}          % JFE-specific formatting of sections, etc.

\DeclareMathOperator*{\plim}{p-lim}

\definecolor{babyblue}{rgb}{0.54, 0.81, 0.94}

\begin{document}

  %!TEX root = ../RCVJ.tex

\setlist{noitemsep}  % Reduce space between list items (itemize, enumerate, etc.)
%\onehalfspacing      % Use 1.5 spacing
% Use endnotes instead of footnotes - redefine \footnote command

\title{Risk of Bitcoin Market: Volatility, Jumps, and Forecasts$^*$
\footnotetext{$^*$ Financial supported by Deutsche Forschungsgemeinschaft via IRTG 1792 \enquote{High Dimensional Nonstationary Time Series}, Humboldt-Universit\"at zu Berlin, the European Union’s Horizon 2020 research and innovation program \enquote{FIN-TECH: A Financial supervision and Technology compliance training programme} under the grant agreement No 825215 (Topic: ICT-35-2018, Type of action: CSA), the European Cooperation in Science \& Technology COST Action grant CA19130 - Fintech and Artificial Intelligence in Finance - Towards a transparent financial industry, the Yushan Scholar Program of Taiwan and the Czech Science Foundation’s grant no. 19-28231X / CAS: XDA 23020303 are greatly acknowledged. All supplementary and Python code can be obtained: \href{https://github.com/QuantLet/RiskofBitcoin_VolaJumpForecast}{\qlet RiskofBitcoin}
}
\footnotetext{$\dagger$ School of Business and Economics, Humboldt-Universit\"at zu Berlin. Spandauer Str. 1, 10178 Berlin, Germany. Email: junjie.hu@hu-berlin.de}
\footnotetext{$\ddagger$ Department of International Business and Risk and Insurance Research Center, National Chengchi University, Taipei City, Taiwan 116. Email: wkuo@nccu.edu.tw}
\footnotetext{$\S$ Blockchain Research Center, Humboldt-Universit\"at zu Berlin, Germany. Wang Yanan Institute for Studies in Economics, Xiamen University, China. Sim Kee Boon Institute for Financial Economics, Singapore Management University, Singapore. Faculty of Mathematics and Physics, Charles University, Czech Republic. National Chiao Tung University, Taiwan. Email: haerdle@wiwi.hu-berlin.de}
}

\author{Junjie Hu$^\dagger$, Weiyu Kuo$^\ddagger$, Wolfgang Karl Härdle$^\S$\\} %

\date{First Draft: Dec. 2019\\ This Draft: Aug. 2021}  % No date for final submission

% Create title page with no page number

\renewcommand{\thefootnote}{\fnsymbol{footnote}}

\singlespacing

\maketitle

\vspace{-.2in}
\begin{abstract}
\noindent

Cryptocurrency, the most controversial and simultaneously the most interesting asset, has attracted many investors and speculators in recent years. The visibly significant market capitalization of cryptos also motivates modern financial instruments such as futures and options. Those will depend on the dynamics, volatility, or even the jumps of cryptos. We provide a comprehensive investigation of the risk dynamics of the Bitcoin Market from a realized volatility perspective. The Bitcoin market is extremely risky in the sense of volatility, entangled jumps, and extensive consecutive jumps, which reflect the major incidents worldwide. Empirical study shows that the lagged realized variance increases the future realized variance, while the jumps, especially positive ones, significantly reduce future realized variance. The out-of-sample forecasting model reveals that, in terms of forecasting accuracy and utility gain, investors interested in the long-term realized variance benefit from explicitly modelling the jumps and signed estimators, which is unnecessary for the short-term realized variance forecast.

\end{abstract}

\medskip



\thispagestyle{empty}

\clearpage

\onehalfspacing
\setcounter{footnote}{0}
\renewcommand{\thefootnote}{\arabic{footnote}}
\setcounter{page}{1}

   %!TEX root = ../RCVJ.tex

\section{Introduction}

% Introduction Part Points:\\
%
% - [Bitcoin market is a segmented and emerging market that draws more and more attention]\\
% - [The importance of Bitcoin market requires more studies]\\
% - [literature Review]
% -- Potential implications for deploying blockchain technology \cite{yermack2017corporate}
% -- Volume predict returns of bitcoin \cite{balcilar2017can}
% -- Change point detection \cite{Thies2018}, structure break of returns and volatility happened frequently
% -- Network perspective of volatility \cite{Yi2018},
% - [Previously, literature more focus on returns of this asset, however, the risk characteristics need more discussion]
% - [From volatility aspect, literature study cryptocurrency using GARCH]\\

Understanding and managing the risk of the cryptocurrency market is crucial for financial investment and the construction of contingent claims.
The popularity of cryptocurrency investment has been rising along with the discussions on blockchain technology applications, for example, the recent Libra from Facebook. However, many of the investors are not informed, or cautious enough about their portfolio risk contributed by cryptocurrency. Different assets may have similar risk characteristics, and the cryptocurrency can be viewed as an outlier characterized by extremely high volatility and more frequent jumps.

Among all the cryptocurrencies, we are motivated to single out the risk of Bitcoin (BTC) for its dominant market share (more than 70\%) and active trading. BTC was first proposed by \cite{nakamoto2008bitcoin} and then initialized in 2009. It is built on blockchain technology which decentralizes and distributes information through networks worldwide. Thus BTC is a naturally decentralized currency as being part of the blockchain. Nevertheless, most of the exchanges where trade BTC and any other cryptocurrencies are regulated, and the data used in this paper is provided by those regulated exchanges.

We are motivated to study the volatility of BTC for the reasons as follows. First is the fact that the BTC price has frequently experienced extreme volatility and jumps since 2013. The Bitcoin market started to draw attention in 2013 when the unit price exceeded \$100. Four years later, in January 2017, the unit price hit \$1000 and reached almost \$20,000  by the end of 2017. The bubble burst in 2018, its price dropped around 80\% from the peak in one year, while it climbed up again in 2019. In our sample period from 2017 to 2020, we observe that the 5-minute logarithmic returns of BTC span from -18.64\% up to 10.30\%. Secondly, the rapid development of Bitcoin and its derivatives market demand studies on the volatility and jump process. Apart from many of the online exchanges offering BTC futures and options, the strictly regulated exchange CME launched futures on BTC in 2017. More and more investors have been entering the BTC market and cause the daily trading volume to rise from around \$100M at the beginning of 2017 to around \$29,000M ~ \$50,000M by the end of 2020\footnote{Data source: www.coinmarketcap.com}. There were some studies that ride on those phenomena. A recent study, \cite{conrad2018long} decompose the volatility into short-term and long-term components by GARCH-MIDAS analysis, and study the volatility correlation between BTC and some other indices, for example, the Baltic dry index. \cite{10.1093/jjfinec/nby013} analyze the jump behavior using the dataset from Mt. Gox exchange in the sample period from 2011 to 2013. And \cite{svcj} attempt to calibrate an option pricing model adapting the high volatility and jump properties. Many other papers focus on the forecasting side, for example, \cite{bitcoin_volatility}.

In this paper, we find that the jump estimator separated from Realized Variance ($RV$) suffers from the consecutive jump phenomenon, which causes the jump estimator biased. $RV$, accounting for intraday information from high-frequency data, is essentially the sum of squared returns over the period(see e.g \cite{andersen2001distributionexchangerate}; \cite{barndorff2002econometric}). The vanilla way to separate jumps from realized variance is by \cite{barndorff2004power}. However, our empirical result shows that this method is impaired by the so-called "consecutive jumps", which is a typical phenomenon in Bitcoin Market. We propose to correct the bias by employing the thresholded jump estimator from \cite{corsi2010threshold}. Two interesting findings on jump risk are insightful. Firstly, despite the extraordinary amount and large size of jumps detected in BTC, the discontinuities do not contribute much to the risk. Moreover, a simple equal-weighted portfolio of BTC from several different exchanges reduces the idiosyncratic jump risk significantly. To further investigate the asymmetric effect, we decompose realized variance into upside risk and downside risk, i.e., realized semi-variance (\cite{barndorff2008measuring}) and then obtain the positive and negative jump components.

The forecasting properties of BTC realized variance is conducted first with 4 HAR-type full-sample forecasting regression and then the 90-day rolling window adaptive forecasting. The full-sample regression shows that the 1-day lagged realized variance estimators and jump estimators increase the future realized variance significantly across the three forecasting horizons $h=1,7,30$, while among all the decomposed jump estimators, only the positive jump estimator carries a negative impact on the future 7-day and 30-day realized variance. Then, the adaptive rolling window forecasting shows that the signed jumps can be a significant predictor of the future realize variance of the longer horizon, e.g., 30-day. 
Also, the coefficients of predictors are evolving systematically, implying the existence of structural breaks, which can be plausible to explain the contradictory finding in the literature. For instance, \cite{andersen2007roughing} find a negative relationship between jumps and future $RV$, and \cite{corsi2010threshold} document that the threshold jump estimator has a significant positive relationship with future $RV$. 

Further analysis on out-of-sample forecasting reveals that the forecasting horizon plays a key role in the decision of whether to model the separate jumps and decomposed signed estimators to forecast BTC realized variance. Despite the previous finding that jumps include extra information, e.g., \cite{nolte2015economic} conclude that modeling jumps explicitly improves the forecasting. We find that in the short forecasting horizon, $h=1$, the HAR model with only lagged $RV$ outperforms all the other models accounted for jumps or signed estimators. As the forecasting horizon gets longer, $h=30$, separating jumps improve forecasting accuracy significantly by the Diebold-Mariano test. Such a finding is further confirmed from an economic point of view by a utility-based framework (\cite{bollerslev2018risk}).

\subsection{Literature Review}

The risk of BTC has been discussed from different angles. Concerning its obvious regulatory risk as an unprecedented "currency" not issued or endorsed by governments, existed literature studies its fundamental risk.
Such as, \cite{yermack2015bitcoin} argues that BTC is rather a speculative investment than a "currency" because of reasons such as its price is too volatile for users, low acceptance from common merchants, etc. \cite{10.1093/jjfinec/nby023} and \cite{gerlach2019dissection} find strong speculative bubble properties in both CRIX (\cite{trimborn2018crix}) and BTC. \cite{griffin2018bitcoin} document possible price manipulations. Due to the lack of fundamental value, \cite{bukovina2016sentiment}; \cite{garcia2015social}; \cite{balcilar2017can} find that the latent drivers of BTC price and volatility could be the sentiment or a series of social signals such as opinions and trading volume. A recent paper of \cite{hardlephenotypic} classify cryptocurrency as a new asset class by its statistical features.
Moreover, the risk of BTC is considered from the aspect of portfolio management. BTC is found to function as a hedging or risk haven asset (\cite{Bouri2017}; \cite{urquhart2019bitcoin}) and it has similar properties like gold under the asymmetric GARCH models (\cite{gronwald2014economics}; \cite{Dyhrberg2016}).  \cite{glaser2014bitcoin} argue that people use BTC not for transactions but as an alternative investment.

We proceed with the article as follows.
In Section \ref{sec:riskestimators}, we briefly describe the realized variance and jump estimators used in this article.
Then, in Section \ref{sec:datasource_pre_analysis}, we present the data we use, followed by a discussion on BTC price processes and summary statistics on (semi-)realized variances and jumps.
Section \ref{sec:forecasting} discusses the construction and comparison of forecasting models, and the forecasts are evaluated under a utility-based framework. Finally, we conclude our findings and remarks in Section \ref{sec:conclusion}.
  %!TEX root = ../RCVJ.tex

\section{Realized Risk: Volatility and Jumps} \label{sec:riskestimators}
This section briefly introduces and justifies the risk estimators employed in the Bitcoin market and the forecasting models. Among many risk estimators in the literature, we choose the realized variance, hereafter $RV$, to evaluate the risk of the Bitcoin market for the following reasons.
First, $RV$ is the most commonly used and examined ex-post risk estimator that does not require information from the derivatives. Second, this estimation, simply accumulating the squared logarithmic price changes, incorporates high-frequency market information compared to other historical volatility estimations. Furthermore, $RV$ enables the estimation of jumps on price, which is one of the main characteristics of the Bitcoin market. The variance estimator originates from the discussion of quadratic variation theory followed by a huge literature on the in-fill asymptotics. The theoretical details and empirical investigations can be referred to \cite{andersen2001distribution, andersen2001distributionexchangerate, barndorff2002estimating,barndorff2004econometric}

\subsection{Realized Variance and Smoothing Variance}

The definition of realized variance $RV_{t+1}$ on a logarithmic asset price process $p(t)$ over one period $(t,t+1]$ is

\begin{equation}\label{eq:def_rv}
 RV_{t, t+1} \stackrel{\operatorname{def}}{=} \sum^{1/\Delta}_{j=1}r^2_{t+j\Delta},
\end{equation}

where $\Delta$ associates with the sampling period, we sample the daily price process every 5-minutes evenly, which means 288 price points per trading day\footnote{24-hours trading in BTC exchanges}, hence $\Delta=1/288$. The logarithmic return $r_{t+j\Delta}$ denotes the $j\mbox{-}th$ observed price change value in day $t$.

Under the classic continuous-time jump diffusion assumptions where price follows \eqref{eq:def_continuous_jump}, by the theory of quadratic variation, $RV$ can be interpreted as the linear combination of a integrated variance component ($IV = \int^{t+1}_{t} \sigma^2(s)ds$) and squared jump component ($J^{2}=\sum_{t<s\leq t+1} \kappa^2(s)$) as the sampling period $\Delta$ goes to zero.

\begin{equation}\label{eq:def_continuous_jump}
 dp(t)=\mu(t)dt+\sigma(t)dW(t)+\kappa(t)dq(t),0\leq t \leq T
\end{equation}

\begin{figure}[!htb]
    \centering
    \begin{minipage}{1\linewidth}
      \includegraphics[width=\textwidth]{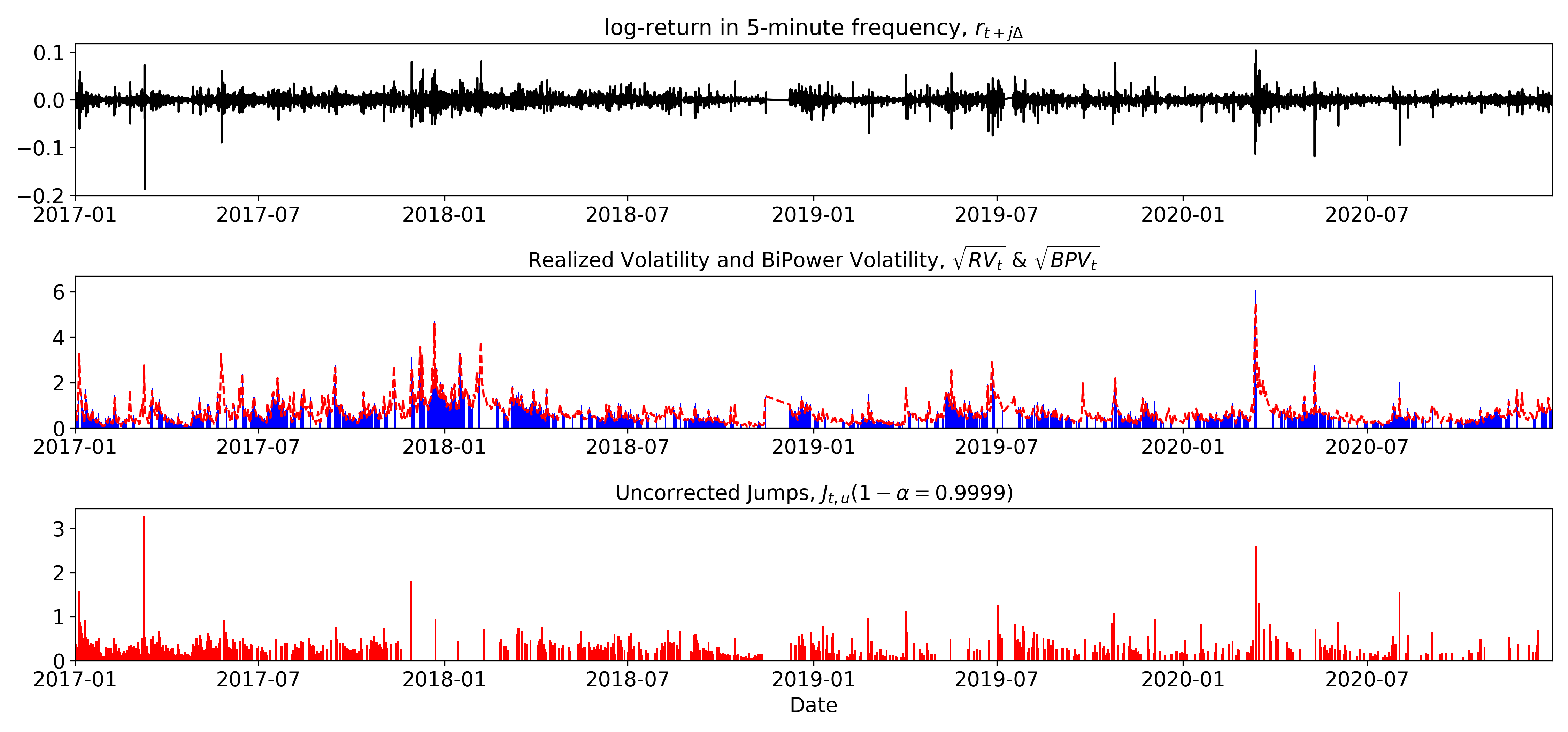}
      % \rule{\linewidth}{10em}
    \end{minipage}
    \caption{
      \textbf{Processes of log return and Realized Risk Measures}
    \newline 
    \small
    Top panel is the log return process ($r_{t+j\Delta}$) of the Bitcoin market in 5-minute frequency. Middle panel contains the realized volatility process ($\sqrt{RV_t}$) in \textcolor{blue}{blue bars}, and the bipower volatility ($\sqrt{BPV_t}$) in \textcolor{red}{red dashed line}. And the bottom panel is the uncorrected separated jump process ($J_{t,u}$) in \textcolor{red}{red bar} at $99.99\%$ confidence level. All processes start from Jan. 1st, 2017 to Dec. 31st, 2020 (UTC$+$0).}\label{fig:lgrt_rvbpv_jump}
  \end{figure}

To separate the squared jump component from the $RV$, many integrated variance estimators were proposed (\cite{barndorff2004power, barndorff2006econometrics, andersen2007roughing}), the primary one is named BiPower Variance estimator, $BPV_{t+1} = \mu_{1}^{-2}\sum_{j=2}^{1/\Delta} |r_{t+j\Delta}| \cdot |r_{t+(j-1)\Delta}|$, where $\mu_1=\sqrt{2/\pi}$. Intuitively, by accumulating the adjacent logarithmic returns, bipower variance intends to smooth the variance estimation and converges to integrated variance as $\Delta$ goes to zero in the presence of infrequent jumps. Then the squared jump component $J^{2}_{u}$ can be isolated by differencing $RV$ and $BPV$, i.e $J^{2}_{t+1,u} \stackrel{\operatorname{def}}{=} \max\left\{RV_{t+1} - BPV_{t+1},0\right\}$\footnote{To guarantee the non-negative of the squared jump component, we simply truncate the negative values. The $u$ in the subscript denoted for the "uncorrected" to distinguish with the "corrected" version in the later section.}. More details about the convergence of realized variance and bipower variance are documented in Appendix \ref{app:converge_prop_rvj}. We show the estimated processes of Bitcoin including risk measures and the uncorrected jump component in Figure \ref{fig:lgrt_rvbpv_jump}, along with the 5-minute log return process.

\subsection{Consecutive Price Jumps in Bitcoin Market}\label{sec:tj}

Separating the jump component in BTC by the BiPower Variance suffers from the consecutive jump and thus causes bias in the jump estimator.
In this subsection, we propose to tackle this problem by employing the threshold BiPower Variance.

\begin{figure}[!htb]
    \centering
    \begin{minipage}{1\linewidth}
      \includegraphics[width=\textwidth]{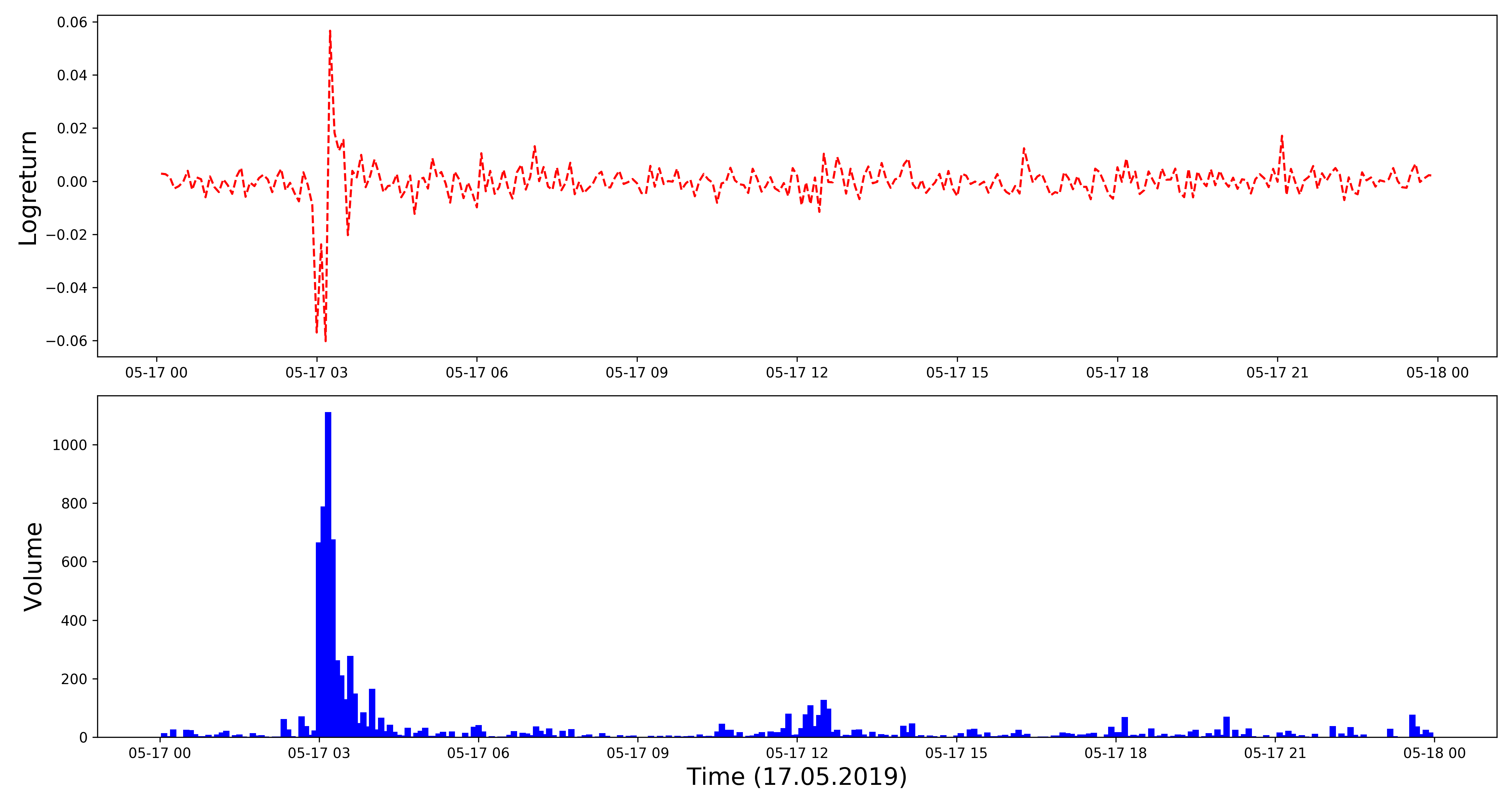}
      % \rule{\linewidth}{10em}
    \end{minipage}
    \caption{
      \textbf{A Demonstration of Consecutive Jumps}
    \newline 
    \small
    Bipower variance is biased to consecutive jumps: A intrady trajectory of Bitcoin price on 17th May 2019.
    $RV=6.09$, $BPV=6.44$ imply that no jump has happened on this day.
    The upper panel shows the logarithmic return process and the bottom panel is the corresponding trading volume. Both panels are in 5-minute frequency.}
    \label{fig:consecutive_jump}
  \end{figure}

In Figure \ref{fig:lgrt_rvbpv_jump}, one can find that most of the spiking log returns are identified as jumps. However, one can also see that jumps are not properly identified in certain periods, such as Dec.2017 to Jan.2018. We argue that such bias can be attributed to the phenomenon of consecutive jumps. The consecutive jumps were not addressed empirically enough in the existing literature; however, they are significant in the Bitcoin market. For instance, as shown in the upper panel of Figure \ref{fig:consecutive_jump}, in the sample of May 17th, 2019, the price process apparently contains a series of jumps around 3 AM (UTC+0). One can compute $RV$ and BiPower variance, $RV_t = 6.09$, $BPV_{t} = 6.44$, which implies that no jump can be detected on that day. It is intuitive to understand how the "false negative" error is caused by consecutive jumps that happens during the period when the market has strong disagreements, which can be evidenced by the simultaneous surge in trading volume as shown in the bottom panel of Figure \ref{fig:consecutive_jump}.

BiPower variance is effective in smoothing a variation process when the jumps are relatively small in size and less frequent, while in the case of the Bitcoin market where investors/speculators are reacting dramatically to any new information or market sentiment, it causes problems. Threshold BiPower Variance ($TBPV$) estimator provided by \cite{corsi2010threshold}, which is a corrected version to relieve the "double-sword" bias in the estimator proposed in \cite{mancini2009non}\footnote{More details are documented in the Appendix \ref{app:converge_prop_tmpv}}, is designed to tackle this problem. Essentially, threshold bipower variance replaces the simple log return with the corrected return, $r^{C}_{1}(r_{t+j\Delta},\theta_{t+j\Delta})$ that takes the expected return $r^e(\theta_{t+j\Delta})$\footnote{Under the assumption that $r_{t+j\Delta}\thicksim \mathcal{N}(0,\sigma^2)$} when the squared return $r^2_{t+j\Delta}$ is larger than a certain threshold value $\theta_{t+j\Delta}$, otherwise the absolute return $|r_{t+j\Delta}|$. Formally, threshold bipower variance can be written as

\begin{equation}
 TBPV_{t+1}=\mu_1^{-2}\cdot\sum_{j=2}^{1/\Delta}r^{C}_1(r_{t+j\Delta},\theta_{t+j\Delta})r^{C}_1(r_{t+(j-1)\Delta},\theta_{t+(j-1)\Delta}),
\end{equation}

where $\mu_1 = \sqrt{2/\pi}$. More details for the general form of the corrected $\eta\mbox{-}th$ power return $r^{C}_{\eta}(r_{t+j\Delta},\theta)$ are documented in Appendix \ref{app_conditional_return}. Note that the threshold value $\theta_{t+j\Delta}$ is a function of local variance $\sigma^2_{t+j\Delta}$, i.e $\theta_{t+j\Delta} = c^2\cdot\sigma_{t+j\Delta}^2$. The given constant $c$ controls how far a return deviates from zero should be classified as a jump, i.e as $c$ goes larger, $TMPV$ truncates fewer values. In our main empirical results, we follow the literature and choose $c=3$. One intuitive way to consider the choice of $c=3$ is the $3\mbox{-}\sigma$ rule of thumb. We also test $c=\{2, 2.5, 3.5, 4\}$, the results of detected jumps are robust. An unbiased estimator $\widehat{V}_{t+j\Delta}$ in \cite{fan2008nonlinear} is implemented to estimate the local variance $\sigma^2_{t+j\Delta}$ (\cite{corsi2010threshold}), see Appendix \ref{app_local_V} for more details.

\begin{figure}[!htb]
\centering
\begin{minipage}{1\linewidth}
    \includegraphics[width=\textwidth]{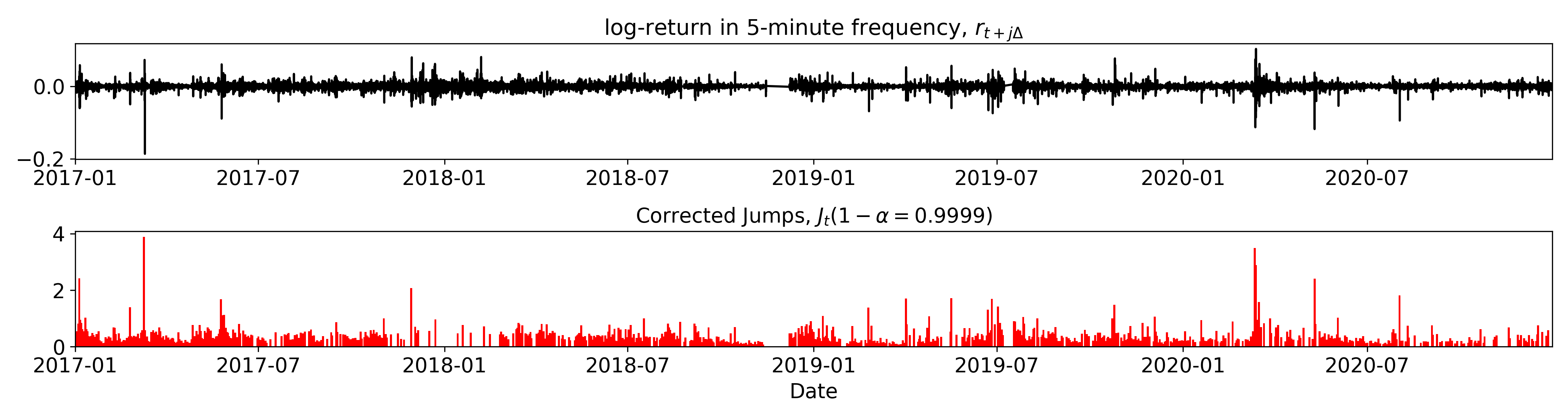}
    % \rule{\linewidth}{10em}
\end{minipage}
\caption{
    \textbf{Log return Process and the Corrected Significant Jump Process}
\newline 
\small
The log return process ($r_{t+j\Delta}$) in the top panel and the corresponding jump process ($J_t(1-\alpha)$) in the bottom panel. The jump process is corrected by applying threshold bipower variation method (\cite{corsi2010threshold}) to overcome the consecutive jumps phenomenon, and the jumps are significant at $0.01\%$ level.}
\label{fig:lgrt_sig_cjump}
\end{figure}

The thresholded jump test $t\mbox{-}z$ is developed based on the ratio statistics (\cite{10.1093/jjfinec/nbi025, andersen2007roughing, corsi2010threshold}). 
Under a series assumptions, for the null hypothesis that no jump exists, $t\mbox{-}z$ converges to standard normal distribution as $\Delta$ goes to 0, i.e $t\mbox{-}z \stackrel{\mathcal{L}}{\rightarrow}\mathcal{N}(0,1)$. 
More details for the jump test statistics are documented in Appendix \ref{app:converge_prop_tmpv}.
Then, the corrected squared jump $J^{2}(1-\alpha)$ at $\alpha$ significance level is

\begin{equation}\label{eq:tj_sig}
 J^{2}_{t+1}(1-\alpha)\stackrel{\operatorname{def}}{=}\max\left\{{RV_{t+1}-TBPV_{t+1}},0\right\}\cdot\mathbf{I}\left\{t\mbox{-}z_{t+1}>\Phi^{-1}_{1-\alpha}\right\}.
\end{equation}

Immediately, we can see that the corrected jumps $J_t(1-\alpha=0.9999)$ in Figure \ref{fig:lgrt_sig_cjump} are much larger in size and much more in quantity compared with the uncorrected version in Figure \ref{fig:lgrt_rvbpv_jump}. Many of the spiking log returns that were missed in the previous jump estimator are now identified. The plot is direct evidence to justify the use of the threshold BiPower variation estimator.
Finally, we define the continuous component by differencing $RV$ and $J^{2}$, i.e $C_{t+1}\stackrel{\operatorname{def}}{=}RV_{t+1}-J^{2}_{t+1}$ to guarantee that $RV=C+J^{2}$.

\subsection{Signs of the Detected Jumps in Bitcoin Market}\label{sec:signed_jump}

\begin{figure}[!htb]
  \centering
  \begin{minipage}{1\linewidth}
      \includegraphics[width=\textwidth]{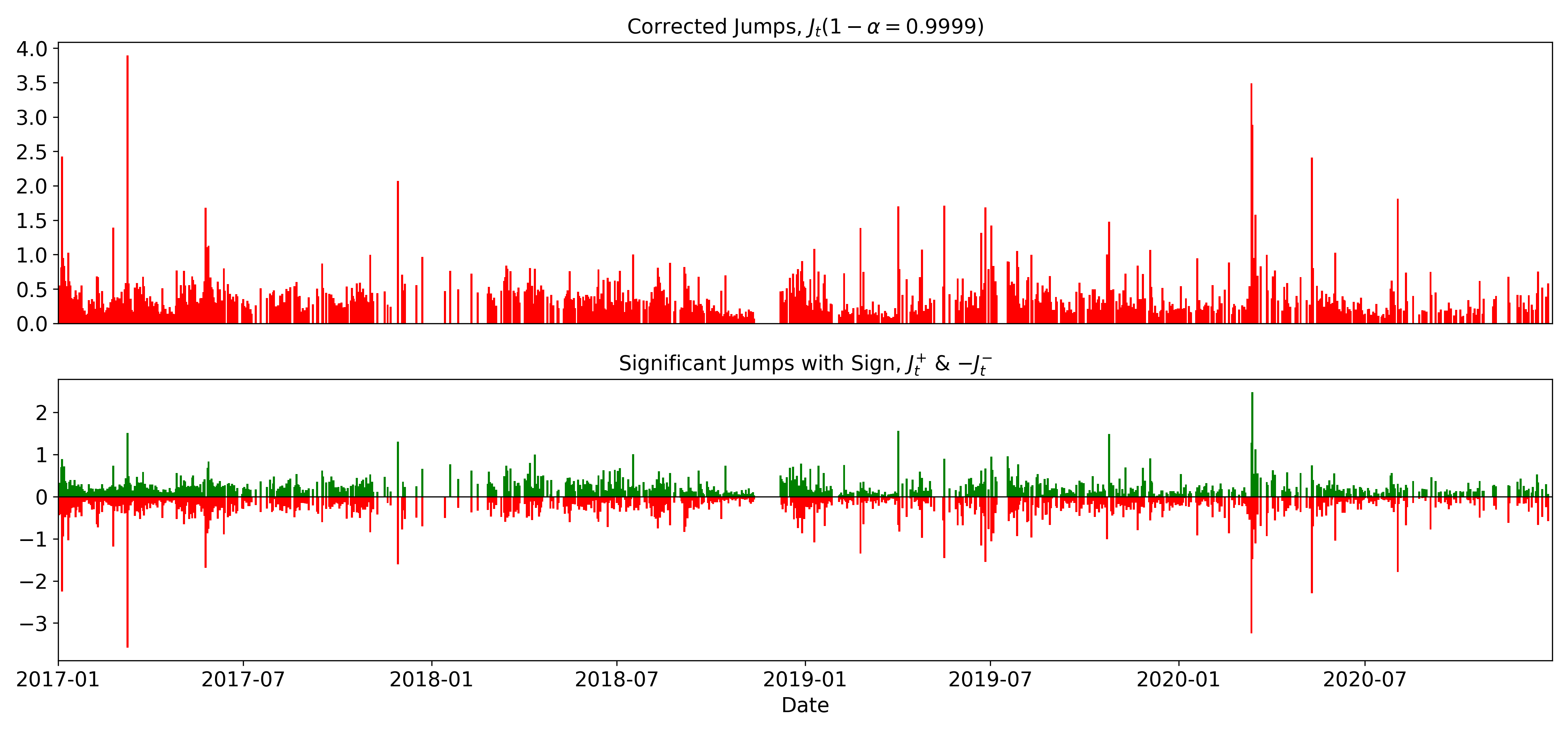}
      % \rule{\linewidth}{10em}
  \end{minipage}
  \caption{
      \textbf{Detected Jumps and Signed Jumps}
  \newline 
  \small
  The detected jump process, $J_t(1-\alpha=0.9999)$, in the top panel and the corresponding signed jump process, $J^{+}_{t}$ \& $-J^{-}_t$ in the lower panel. The signs of jumps are only identified on the days with significant jumps, and both $J^{+}$ and $J^{-}$ are non-negative in value.}
\label{fig:lgrt_pos_neg_jump}
\end{figure}

Many existing models suggest the asymmetric responses to upward and downward risk on asset returns,
and investors are keen to understand the after-effects of extreme market performance.
Unfortunately, the jumps estimator in the previous section does not account for the sign of jumps and is non-negative in theory.
We employ the realized semi-variance provided by \cite{barndorff2008measuring} to obtain the positive and negative jumps.

Realized semi-variance provides a way to separate positive and negative jumps from the realized variance process, 

\begin{equation}\label{eq:sv}
    RSV^{+}_{t+1} = \sum^{1/\Delta}_{j=1}r_{t+j\Delta}^2\cdot \mathbf{I}\{r_{t+j\Delta}>0\}
\end{equation}
   
For simplicity, here we take the positive realized semi-variance for illustration, and the negative estimator follows. Recall that under the continuous jump diffusion model assumption, the jump component of quadratic variation $QV_{t+1}$ is the accumulated sum of squared infinitesimal changes $\Delta p_s=p_s - p_{s_-}$, i.e $\sum_{t<s\leq t+1} (\Delta p)^2(s)$. $RSV^{+}$ shown in Equation \eqref{eq:sv} is essentially the sum of the squared positive logarithmic returns. It is straightforward that $RV$ can be decomposed into $RSV^+$ and $RSV^-$ completely, i.e $RV = RSV^+ + RSV^-$, for both finite sample and large sample cases.
As sampling frequency $1/\Delta$ goes infinite, the limiting behavior under infill asymptotics shows that $RSV^{+}$ converges to one-half of the integrated variance, $\frac{1}{2}\int^{t+1}_{t} \sigma^2(s)ds$, and the sum of squared positive jumps, $\sum_{t<s\leq t+1} (\Delta p_s)^2\cdot\mathbf{I}\{\Delta p_s>0\}$. Spontaneously, the positive jump component is defined as,

\begin{equation}\label{eq:signed_j}
    J^{2(+)}_{t+1} (1-\alpha) \stackrel{\operatorname{def}}{=} \operatorname{max}\left\{RSV^{+}_{t+1} - \frac{1}{2}TBPV_{t+1},0\right\}\cdot\mathbf{I}\left\{ J^{2}_{t+1}(1-\alpha) > 0 \right\}.
\end{equation}
   
Note that here we propose to only identify the signed jumps on the days with significant jumps, i.e., $J^{2}(1-\alpha)>0$, for the lack of test of significance on the signed jumps. With the properties of $TBPV$ in Appendix \ref{app:converge_prop_tmpv}, one can easily separate the squared positive jumps, i.e., the positive jump component.

\subsection{HAR Models}\label{subsec:har_model}

This subsection introduces four HAR-type models, abbreviated as HAR, RVJ, RSV, RSVSJ, that are used in the forecasting analysis section.
The four models are aimed to investigate the response of realized risk to the lagged risk estimators and the necessity of modelling the jumps components in the Bitcoin market in terms of risk management.

The temporal dependence structure of the realized variance process is crucial in forecasting.
Many of the studies use different ARCH, ARMA, and stochastic volatility models to capture the temporal dependency or long-memory effect.
Other than those complicated models, a simple linear model named the Heterogenous AutoRegression model, HAR, is proposed in \cite{corsi2009simple}.
The advantages of HAR can be illustrated threefold.
First, it is a parameter parsimonious volatility regression model that can be constructed easily with different lagged regressors.
Second, it captures the strong temporal dependency and shows good forecasting performance comparing with those complicated models.
Finally, HAR can be extended by using any other relevant estimators, for example, the jump components.
Such extensibility allows one to investigate a wide range of effects on $RV$.

The basic log-log HAR model is formulated as,

\begin{equation}\label{eq:har_def}
    \log(RV_{t,t+h})=\alpha+ \sum_{l={1,7,30}} \log(RV_{t-l:t})\beta_{l} +\varepsilon_{t,t+h},
    % \log(RV_{t,t+h})=\alpha+ \log(RV_{t-1:t}\beta_{D}) + \log(RV_{t-7:t}\beta_{W}) + \log(RV_{t-30:t}\beta_{M}) +\varepsilon_{t,t+h},
\end{equation}
  
where $RV$ variables are in logarithmic form, and the forecasting horizon $h=\{1,7,30\}$.
The explanatory variables are $RV$ aggregated over the past 1-day, 7-day, and 30-day, as the Bitcoin market is trading 24/7.

By accounting for the corrected jump component proposed in Section \ref{sec:tj}, RVJ model essentially add multiple lagged jump estimators, $J_{t-1:t}$, $J_{t-7:t}$, and $J_{t-30:t}$, on the RHS of Equation \eqref{eq:har_def},

\begin{align}\label{eq:rvj_def}
    \log(RV_{t,t+h})=\alpha+ \sum_{l=1,7,30} \log(RV_{t-l:t})\beta_{l} + \sum_{l=1,7,30} \log(J_{t-l:t} + 1)\gamma_{l} +\varepsilon_{t,t+h},
\end{align}

Note that here the logarithmic transform ensures that the jump estimators to be positive.
Furthermore, with $RV$ decomposed into realized semi-variance estimators, the RSV model accounts for the asymmetic effect from the positive risk and negative risk,

\begin{align}\label{eq:rsv_def}
    \log(RV_{t,t+h})=\alpha + \sum_{l=1,7,30} \log(RSV^{+}_{t-l:t})\beta_{l}^{+} + \sum_{l=1,7,30} \log(RSV^{-}_{t-l:t})\beta_{l}^{-} +\varepsilon_{t,t+h},
\end{align}

With analogous arguments, we can formulate the RSVSJ model by extending the RSV with positive/negative jumps,

\begin{align}\label{eq:rsvj_def}
    \log(RV_{t,t+h})=\alpha &+ \sum_{l=1,7,30} \log(RSV^{+}_{t-l:t})\beta_{l}^{+} + \sum_{l=1,7,30} \log(RSV^{-}_{t-l:t})\beta_{l}^{-} \\
    &+ \sum_{l=1,7,30} \log(J^{+}_{t-l:t} + 1)\gamma_{l}^{+} + \sum_{l=1,7,30} \log(J^{-}_{t-l:t} + 1)\gamma_{l}^{-} +\varepsilon_{t,t+h},
\end{align}

All coefficients in the four models are estimated by OLS.
To adjust the possible serial correlation and heteroskedasticity of the error term, we use the Newey-West covariance matrix estimator with 7, 14, and 60 lags for daily, weekly and monthly forecast horizon, respectively.
Note that the aggregated estimators, e.g $RV_{t-l:t}$, are averaged over  period $(t-l:t]$. The averaging method not only has incorporated information over the period but also ensures estimates having the same scale. Here we employ the daily ($l=1$), weekly ($l=7$), and monthly ($l=30$) lagged estimators. Those three stepwise estimators capture the footprint of $RV's$ long memory property. All the jump estimators are based on $\alpha=0.9999$ and $c_{\theta}=3$.

   % !TEX root = ../RCVJ.tex

\section{Empirical Analysis on Realized Risk and Jumps}\label{sec:datasource_pre_analysis}

\subsection{Data}

In this section, based on the price sample from an online exchange, we conduct an empirical analysis on the realized risk and corrected jumps introduced in the previous section. There are 270\footnote{Until Dec 2020, https://coin.market/exchanges-info.php} online exchanges trading various types of cryptocurrencies, and each of the coins is traded in different exchanges globally. Among all crypto markets, the Bitcoin market is dominant with more than 70\% market share, which motivates us to focus only on the risk of BTC in this paper.

We obtain the price data from an online exchange named Gemini\footnote{https://gemini.com/about/}, which is one of the largest digital exchanges regulated by the New York State Department of Financial Services (NYDFS). Furthermore, we would like to relieve the concerns on price jump of idiosyncratic exchange by demonstrating a synthetic process constructed by averaging prices from three actively trading exchanges, Poloniex, Bittrex, and Bitfinex\footnote{Poloniex and Bittrex are U.S based companies, and Bitfinex located in Hong Kong}.

After cleaning and removing the trading days with missing data points, the dataset starting from Jan. 2017 until Dec. 2020 contains 1385 trading days\footnote{Samples from 15th Nov. 2018 to 6th Dec. 2018. are removed for the missing data problem} with a 5-minute sampling frequency in each day. The Bitcoin market is trading all day and all year globally akin to the foreign exchange market, which prompts the issue of removing the illiquid periods, such as weekends/holidays and inactive trading hours. In this paper, we do not remove any trading days because many non-institutional investors also trade during non-business days in the cryptocurrency market. The time frame is synchronized to the UTC$+0$ time zone.

Following most of the empirical literature such as \cite{andersen2001distribution, andersen2007roughing}, we adopt the 5-minute high-frequency sampling strategy, i.e., taking the transaction prices closest to the end of each 5-minute interval to calculate 5-minute returns. Hence, we have 288 samples each day. Higher sampling frequency captures more market information but also suffers more from microstructure noise. It is a trade-off depending on the trading activity of the target market. The trading volume has surged in the Bitcoin market since 2017, however still much lower than most of the traditional financial markets. We argue not to alienate the Bitcoin market from other markets in terms of sampling frequency. A series of literature has discussed the sampling frequency issues that the microstructure noise impairs the realized variance estimator.  \cite{ait2005often, BANDI2008} attempt to derive optimal sampling frequency by explicitly assuming noise structure, \cite{Zhang2005, zhang2006efficient} document the efficient estimator by subsampling schemes, and kernel methods are introduced to handle the noise (\cite{barndorff2008designing, Hansen2006}). \cite{liu2015does} test the estimators constructed with different sampling frequencies and find no evidence against the 5-minute sampling strategy.

\subsection{Realized Volatility in Bitcoin Market}\label{sec:rv_btc}

BTC manifests itself with extremely high risk immediately with its sample mean of $RV$ shown in Table \ref{tab:summary_des_rvs}, where one can see the sample mean of annualized daily realized variance of BTC is 0.86, which is much larger compared with other traditional markets. For instance, the Shanghai SE Composite Index (SSEC) has annualized daily realized variance of 0.23. \footnote{Datasource from Realized Library, Oxford-Man Institute of Quantitative Finance. The trading hour bias is corrected by accounting for the overnight price change (\cite{bollerslev2018risk}) to allow the two $RV$ estimators to be comparable. More realized variance of global indices are detailed in Table \ref{tab:des_indices_btc} of Appendix \ref{app:risk_jump_data}.}
% C is the estimation of sigma square

As reported in Columns (1)-(3) of Table \ref{tab:summary_des_rvs}, the distribution of the three original risk measures, $RV$, $RSV^+$, and $RVS^-$ are right-skewed with heavy fat tails. Take realized variance $RV$ in Column (1) for example. It has a standard deviation of $1.88$, which is much higher than its mean value of $0.86$, a positive skewness of $9.12$ implying right-skewed distribution, and a high excessive kurtosis of $126.82$ indicating a fat tail. It is similar for the two realized semi-variance.

By comparing the summary statistics of continuous component $C$ reported in Column (7) with realized variance, one can see that a large proportion of risk coming from jumps is peeled from $RV$. The continuous component has a smaller sample mean of $0.70$ compared to $0.86$ in $RV$ and a much smaller maximum value of $28.60$ compared with $36.93$ in $RV$. However, the continuous component is still dominant in the realized risk measure.

\begin{table}[!htb]
  \small
  \setlength{\tabcolsep}{0pt}
  \begin{threeparttable}
    \caption{Summary Statistics of Bitcoin Realized Risk Measures}\label{tab:summary_des_rvs}
    \begin{tabular*}{\linewidth}{@{\extracolsep{\fill}}>{\itshape}
      l *{7}{S[table-format=-1.4,table-number-alignment = center]}
      } 
        \toprule
        \toprule
        \multicolumn{1}{c}{}
        & \multicolumn{1}{c}{$RV$}
        & \multicolumn{1}{c}{$RSV^+$}
        & \multicolumn{1}{c}{$RSV^-$}
        & \multicolumn{1}{c}{$\log(RV)$}
        & \multicolumn{1}{c}{$\log(RSV^+)$}
        & \multicolumn{1}{c}{$\log(RSV^-)$}
        & \multicolumn{1}{c}{$C$}
        \\
        % \cline{1-2}
        \midrule
        {}  & \multicolumn{1}{c}{(1)}  & \multicolumn{1}{c}{(2)}  & \multicolumn{1}{c}{(3)}   
        & \multicolumn{1}{c}{(4)}   & \multicolumn{1}{c}{(5)}  & \multicolumn{1}{c}{(6)}     
        & \multicolumn{1}{c}{(7)}   \\[0.15cm] 
      mean     & 0.86   & 0.41   & 0.45    & -0.97 & -1.71 & -1.68 & 0.70 \\
      std      & 1.88   & 0.92   & 1.02    & 1.24  & 1.24  & 1.28  & 1.55 \\
      min      & 0.76\% & 0.40\% & 0.36\%  & -4.88 & -5.51 & -5.63 & 0.18\% \\
      5\%      & 0.05   & 0.02   & 0.02    & -2.96 & -3.74 & -3.71 & 0.03 \\
      50\%     & 0.37   & 0.17   & 0.18    & -1.01 & -1.77 & -1.73 & 0.26 \\
      95\%     & 2.83   & 1.38   & 1.44    & 1.04  & 0.32  & 0.36  & 2.55 \\
      max      & 36.93  & 20.44  & 16.50   & 3.61  & 3.02  & 2.80  & 28.60 \\
      skewness & 9.12   & 10.78  & 8.77    & 0.14  & 0.15  & 0.16  & 8.44 \\
      kurtosis & 126.82 & 185.59 & 106.87  & 0.13  & 0.08  & 0.07  & 110.56 \\
      acf(1)   & 0.45   & 0.42   & 0.42    & 0.74  & 0.73  & 0.71  & 0.47 \\
      acf(7)   & 0.13   & 0.14   & 0.11    & 0.48  & 0.50  & 0.46  & 0.19 \\
      acf(30)  & 0.05   & 0.05   & 0.05    & 0.25  & 0.26  & 0.24  & 0.10 \\
      acf(100) & 0.60\% & 0.01   & -0.13\% & 0.04  & 0.04  & 0.04  & 0.02 \\
      \emph{ADF} & -4.80\threeS & -4.66\threeS & -7.15\threeS & -6.70\threeS & -5.03\threeS & -5.30\threeS & -4.04\threeS \\
        \bottomrule
        \bottomrule
    \end{tabular*}
  \begin{tablenotes}[flushleft]
   \setlength\labelsep{0pt} 
  \linespread{1}\small
  \item Columns (1)-(3) are realized variance, positive realized semi-variance, negative realized semi-variance, followed by the logarithmic form of those three estimators in Columns (4)-(6).
  Column (7) is the continuous component defined as the difference between realized variance and significant corrected jump.
  The first six-row contains the sample mean, standard deviation, minimum, percentiles, maximum, skewness, and excess kurtosis, followed by four autocorrelation functions with 1-day, 7-day, 30-day, and 100-day lags. 
  The last row reports the Augmented Dickey-Fuller test with three significance levels. \threeS: 1\% significance, \twoS: 5\% significance, \oneS: 10\% significance.
\end{tablenotes}
  \end{threeparttable}
\end{table}

\begin{figure}[!htb]
  \begin{subfigure}[normal]{0.5\textwidth}
    \includegraphics[width=\textwidth]{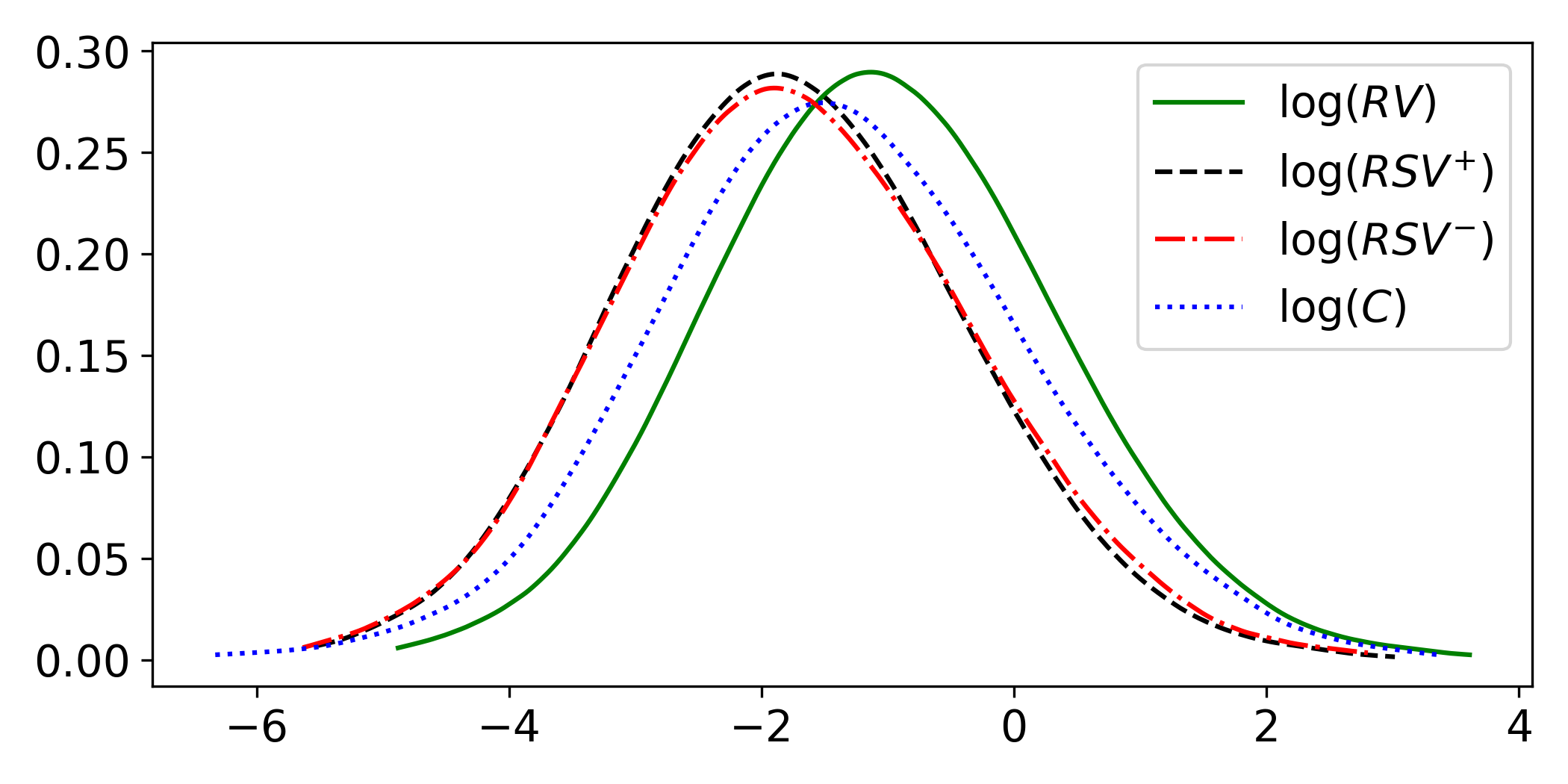}
    \caption{
      \textbf{KDE}
      % \newline 
      % \small
      }\label{fig:kde_rvs_maincoin}
  \end{subfigure}
  \begin{subfigure}[normal]{0.5\textwidth}
    \includegraphics[width=\textwidth]{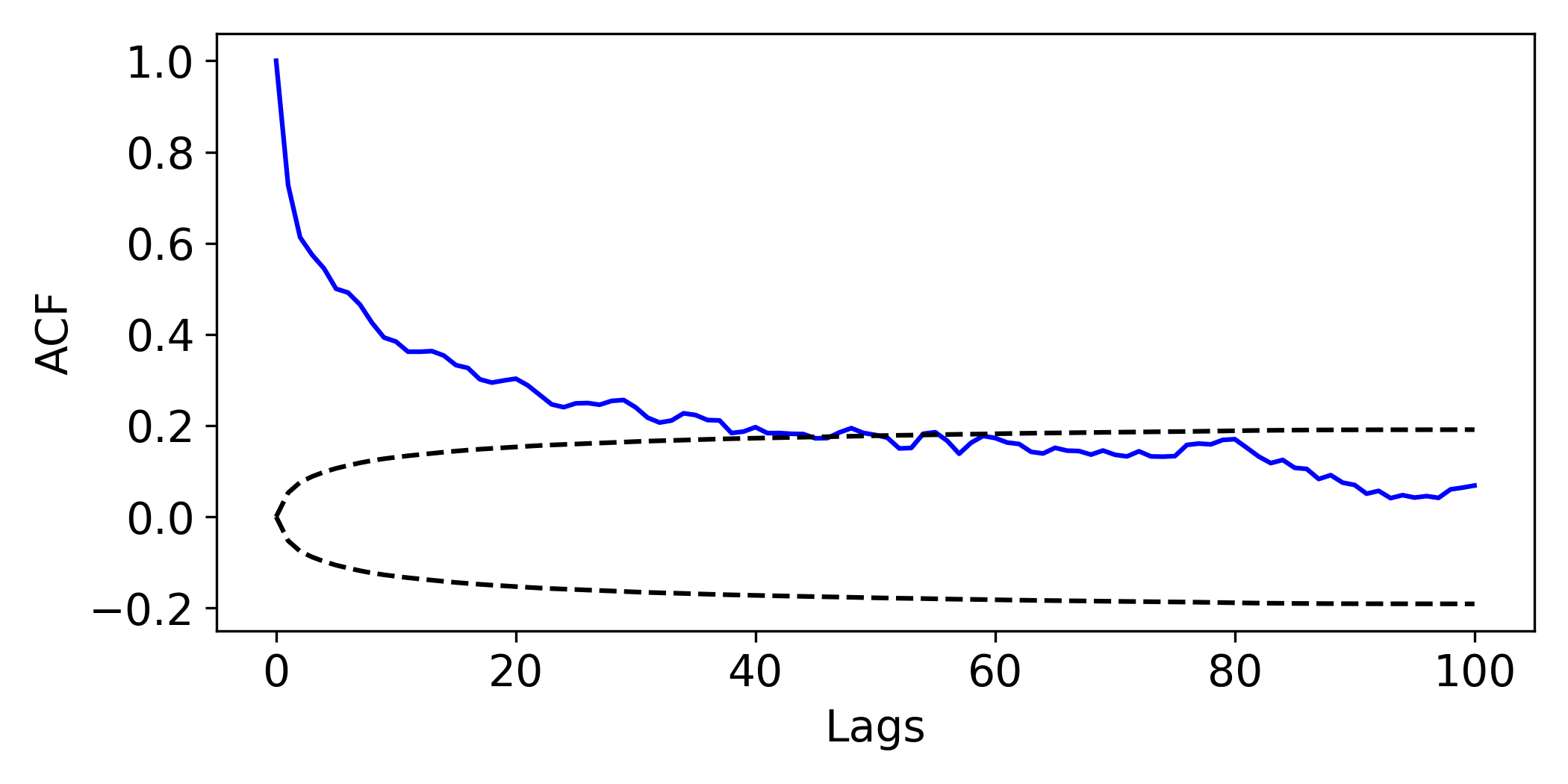}
    \caption{
      \textbf{ACF}
    %   \newline 
    %   \small
      }\label{fig:acf_rvs_maincoin}
  \end{subfigure}
  \caption{
      % \textbf{Kernel Density Estimation and AutoCorrelation Function}
      % \newline 
      \small
      Figure (a) is the Epanechnikov kernel density estimations (bandwidth=1.5) on the logarithmic annualized unconditional daily risk measures including \textcolor{green}{Realized Variance($RV$)}, Realized Semi-Variances (\textcolor{black}{($RSV^+$)}, \textcolor{red}{$RSV^-$}), and the continuous component (\textcolor{blue}{$C$}).
      Figure (b) is the \textcolor{blue}{autocorrelation function (ACF)} of $\log(RV)$ with 95\% confidence band in dashed line.
      }
\end{figure}

% This subsection discusses the dynamics of variance estimates in three aspects, including descriptive statistics, empirical distribution, and series autocorrelation.

% Log-normal distribution of RV, slightly positive skew of the log form
% Persistent temporal dependency
% Stationarity
After taking the logarithmic form, those risk measures become very close to the normal distribution and more persistent, which motivates us to focus on the logarithmic form in the later forecasting studies. It is a stylised fact that $RV$ is characterized by log-normal distribution and strong temporal dependency (\cite{andersen2001distribution}). The skewness and excess kurtosis of all the three realized risk measures in the logarithmic form are close to zero as described in Columns (4)-(6) of Table \ref{tab:summary_des_rvs}. For example, realized variance in the logarithmic form $\log(RV)$ reported in Column (4) has a skewness of $0.13$ and excessive kurtosis of $0.08$ compared to the highly right-skewed and long-tailed squared form in Column (1).
The Epanechnikov kernel density estimation (bandwidth=$1.5$) also shows that those three risk measures in the logarithmic form are close to the normal distribution as shown in Figure \ref{fig:kde_rvs_maincoin}. We also observe strong and persistent temporal dependency of the risk measures. The autocorrelation function values in Table \ref{tab:summary_des_rvs} are significantly large across all the risk measures, especially the measures in logarithmic form, and Figure \ref{fig:acf_rvs_maincoin} depicts the hyperbolic shape of the ACF of $\log(RV)$, the temporal dependency is still significant up to 40 lags.
The long-memory characteristic of $\log(RV)$ in the Bitcoin market motivates the HAR-type models in section \ref{sec:forecasting} to account for daily, weekly, and monthly lagged predictors. All the risk measures are stationary suggested by the ADF test at $1\%$ significance level.

\subsection{Jump Process and Bitcoin Market Incidents}\label{jump_emp}

% Recall that we mainly construct three jump estimators, the corrected thresholded jump $J(1-\alpha)$, and the signed thresholded jump $J^{+(-)}$ in the previous section.
% We first explain the descriptive statistics of the three jump estimators and then show the detected jump processes.
% Only the non-zero detected jumps are reported in Table \ref{tab:summary_des_jumps} and estimator $J(1-\alpha)$ is evaluated at 99.99\% confidence level.

% More jumps in the Cryptocurrency market
Jumps appear far more frequently in the Bitcoin market than in any other developed markets by comparing our results with the literature using similar approaches. As reported in Column (2) of Table \ref{tab:summary_des_jumps}, 64\% of the 1385 days are detected with more than one jump by the corrected jump estimator at $\alpha=99.99\%$ confidence level. The literature shows that in the most liquid six stocks of S\&P500, only 8.3\% of the 1256 trading days are entangled with jumps by the corrected thresholded jump estimator $J(\alpha=99.9\%)$ (\cite{corsi2010threshold}). Up to 8.3\% of all the sample days are detected with jumps in DM/\$ exchange rate market by the uncorrected jump estimator as documented in \cite{andersen2007roughing}.

% To use corrected
By comparing the summary statistics between the corrected and uncorrected jump estimators, the effectiveness of employing the corrected jump estimator is evidenced. First of all, the corrected jump estimator identifies more jumps as reported in Column (1) and (2) of Table \ref{tab:summary_des_jumps}, the corrected jump estimator identifies that $64\%$ of the days are entangled with jumps compared with that of $42\%$ by the uncorrected jump estimator. Secondly, the size of jumps is larger by the corrected estimator. The sample mean of jumps is $0.39$, and the sample maximum is $3.89$ by the corrected jump estimator, while the mean and maximum are $0.33$ and $3.29$ by the uncorrected one. Moreover, the corrected jump estimator is larger by all the percentiles statistics.

Jumps caused by the idiosyncratic risk of exchange do exist and should be noted by investors. There is a concern that many of the jumps are caused by the idiosyncratic risk of the exchange. In other words, some price jumps happen only in a certain exchange but not as reactions to the market. We find evidence to support this concern by investigating a synthetic price process constructed by averaging prices from several exchanges. Table \ref{tab:summary_btcd_jumps} in Appendix \ref{app:risk_jump_data} reports the summary statistics of jump estimators on the synthetic price process, where one can see that jumps identified in the synthetic price process are much fewer compared with the jumps identified in the price process from single exchanges reported in Table \ref{tab:summary_btcd_jumps}. Around $39\%$ of the days are identified with jumps averaging a size of $0.37$ by the corrected jump estimator, while in the single exchange price process, $64\%$ of the days are identified with jumps averaging a size of $0.39$. The rationale of overcoming idiosyncratic jumps in a single exchange by averaging prices from multiple exchanges is similar to the justification of using the index to detect systematic jumps that idiosyncratic jumps are \enquote{wiped out} when the jumps are not reflected in all the exchanges (\cite{A_t_Sahalia_2012}).

\begin{table}[!htb]
  \small
  \setlength{\tabcolsep}{0pt}
  \begin{threeparttable}
    \caption{Summary Statistics For Bitcoin Jump Components}\label{tab:summary_des_jumps}
    \begin{tabular*}{\linewidth}{@{\extracolsep{\fill}}>{\itshape}
      l *{7}{S[table-format=1.3,table-number-alignment = center]}}
       
        \toprule
        \toprule
        \multicolumn{1}{c}{}
        & \multicolumn{1}{c}{$J_{u}(1-\alpha)$}
        & \multicolumn{1}{c}{$J(1-\alpha)$}
        & \multicolumn{1}{c}{$\log(J(1-\alpha)+1)$}
        & \multicolumn{1}{c}{$J^+$}
        & \multicolumn{1}{c}{$J^-$}
        & \multicolumn{1}{c}{$\log(J^+ +1)$}
        & \multicolumn{1}{c}{$\log(J^- +1)$}
        
        \\
        % \cline{1-2}
        \midrule
        {}  & \multicolumn{1}{c}{(1)}  & \multicolumn{1}{c}{(2)}  & \multicolumn{1}{c}{(3)}   
        & \multicolumn{1}{c}{(4)}   & \multicolumn{1}{c}{(5)}  & \multicolumn{1}{c}{(6)}     
        & \multicolumn{1}{c}{(7)}   \\[0.15cm] 

        prop     & 0.42  & 0.64  &      & 0.58   & 0.59   &        & \\
        mean     & 0.33  & 0.39  & 0.31 & 0.25   & 0.29   & 0.22   & 0.23 \\
        std      & 0.26  & 0.32  & 0.18 & 0.21   & 0.29   & 0.14   & 0.17 \\
        min      & 0.06  & 0.06  & 0.06 & 0.12\% & 0.97\% & 0.12\% & 0.97\% \\
        5\%      & 0.10  & 0.12  & 0.11 & 0.06   & 0.06   & 0.06   & 0.06 \\
        50\%     & 0.28  & 0.32  & 0.27 & 0.20   & 0.21   & 0.18   & 0.19 \\
        95\%     & 0.69  & 0.83  & 0.60 & 0.63   & 0.72   & 0.49   & 0.54 \\
        max      & 3.29  & 3.89  & 1.59 & 2.48   & 3.59   & 1.25   & 1.52 \\
        skewness & 4.81  & 4.62  & 2.21 & 3.32   & 4.89   & 1.86   & 2.42 \\
        kurtosis & 41.35 & 35.48 & 8.97 & 21.30  & 39.44  & 6.16   & 10.14 \\
        acf(1)   & 0.15  & 0.22  & 0.23 & 0.15   & 0.14   & 0.15   & 0.15 \\
        acf(7)   & 0.11  & 0.09  & 0.13 & 0.07   & 0.06   & 0.09   & 0.08 \\

        \bottomrule
        \bottomrule
    \end{tabular*}
  \begin{tablenotes}[flushleft]
   \setlength\labelsep{0pt} 
  \linespread{1}\small
  \item Descriptive statistics of three jump estimation of Bitcoin Market.
  Columns from left to right are the corrected thresholded jump estimator $J^{C}(1-\alpha)$ at $\alpha=99.99\%$ confidence level, the thresholded positive jump estimator $J^+$, and the thresholded negative jump estimator $J^-$.
  The threshold constant coefficient $c=3$.
  The first row reports the proportion of non-zero jumps.
  The following rows contain the sample mean, sample deviation, sample minimum, 50\% quantile, and sample maximum.
\end{tablenotes}
  \end{threeparttable}
\end{table}

\begin{figure}[!htb]
  \centering
  \begin{minipage}{1\linewidth}
      \includegraphics[width=\textwidth]{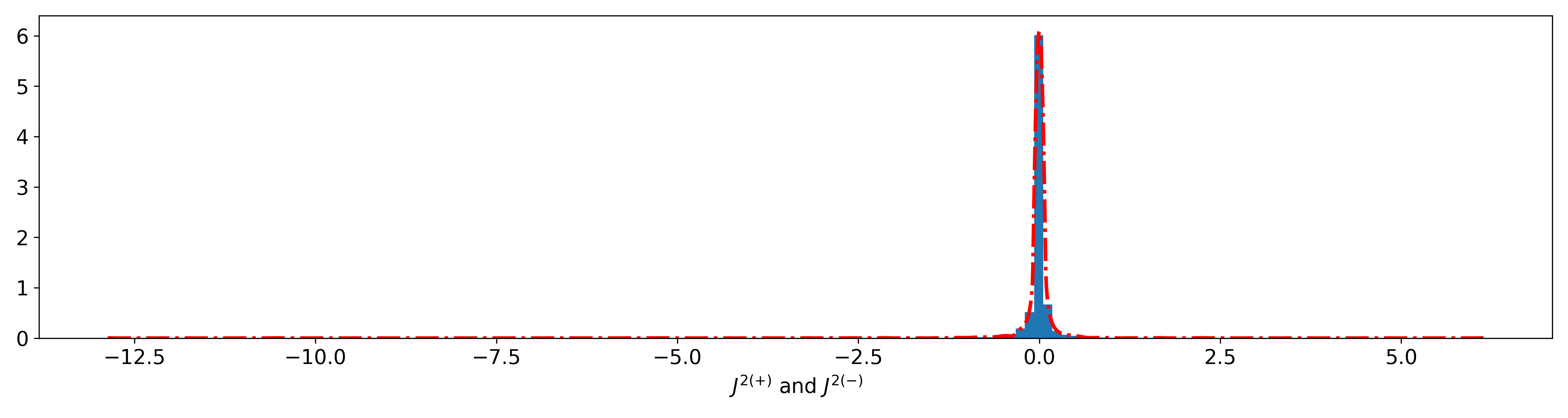}
      % \rule{\linewidth}{10em}
  \end{minipage}
  \caption{
      \textbf{Kernel Density Estimation on Signed Squared Jump Components}
  \newline 
  \small
  An Epanechnikov kernel density estimation on the union of the positive squared jumps and the negative squared jumps, i.e $\{J^{2(+)}\}\cup\{-J^{2(-)}\}$. Note that the negative squared jump component, $J^{2(-)}$, is non-negative.}
  \label{fig:kde_jump_posneg}
\end{figure}

Negative jumps are slightly more often and larger than positive jumps. On the one hand, the negative jumps are larger compared with the positive jumps as reported in Column (4) and (5) in Table \ref{tab:summary_des_jumps}, implying a downward price jump is likely to be more intensive than an upward price jump, even in the upward trending sample period from 2017 to 2020. On the other hand, the proportions of positive and negative jumps detected are similar, which is contrary to the empirical results in \cite{10.1093/jjfinec/nby013} concluding that most jumps in BTC are positive from June 2011 to November 2013. The average size of negative jump detected is $0.29$ each day with a maximum value of $3.59$, while the average size of positive jump is $0.25$ with a maximum size of $2.48$. Figure \ref{fig:kde_jump_posneg} shows the kernel density estimation on the union of squared positive jump estimator and squared negative jump estimator, where we can also see the long tail for the negative jump component. Hence the signed squared jump estimator is left-skewed. Note that the squared negative jump estimator is non-negative, and here we take the negative sign to get a clear look at the distribution of the union of positive and negative jump components.

The logarithmic jump estimators reported in Columns (3), (6), and (7) are employed in the forecasting analysis as the logarithmic form reduces the skewness and kurtosis, hence better time-series properties.

% On the one hand, from the second figure in each panel of positive and negative jump processes, one can see that the jumps appear more frequently comparing the first figure of such panel, e.g., for BTC-D, 76\% of the sample days are detected with positive jumps; however, only 39\% of them are categorized as jump days (See Table \ref{tab:summary_des_jumps}).
% On the other hand, the decomposition of $J(1-\alpha)$ into signed jump estimators is clean in terms of jump intensity.
% The average intensity of jump components decomposition is approximate $J(1-\alpha) \approx J^+ + J^-$.
% This conflict is partially caused by the lack of significance test on signed jumps $J^{+/-}$.
% Also, The size and quantity of negative jumps $J^-$ are almost equal to that of positive jump $J^+$ which implies that a jump is not necessary a crash event in the sample period used in this article.

\begin{figure}[!htb]
  \centering
  \begin{minipage}{1\linewidth}
      \includegraphics[width=\textwidth]{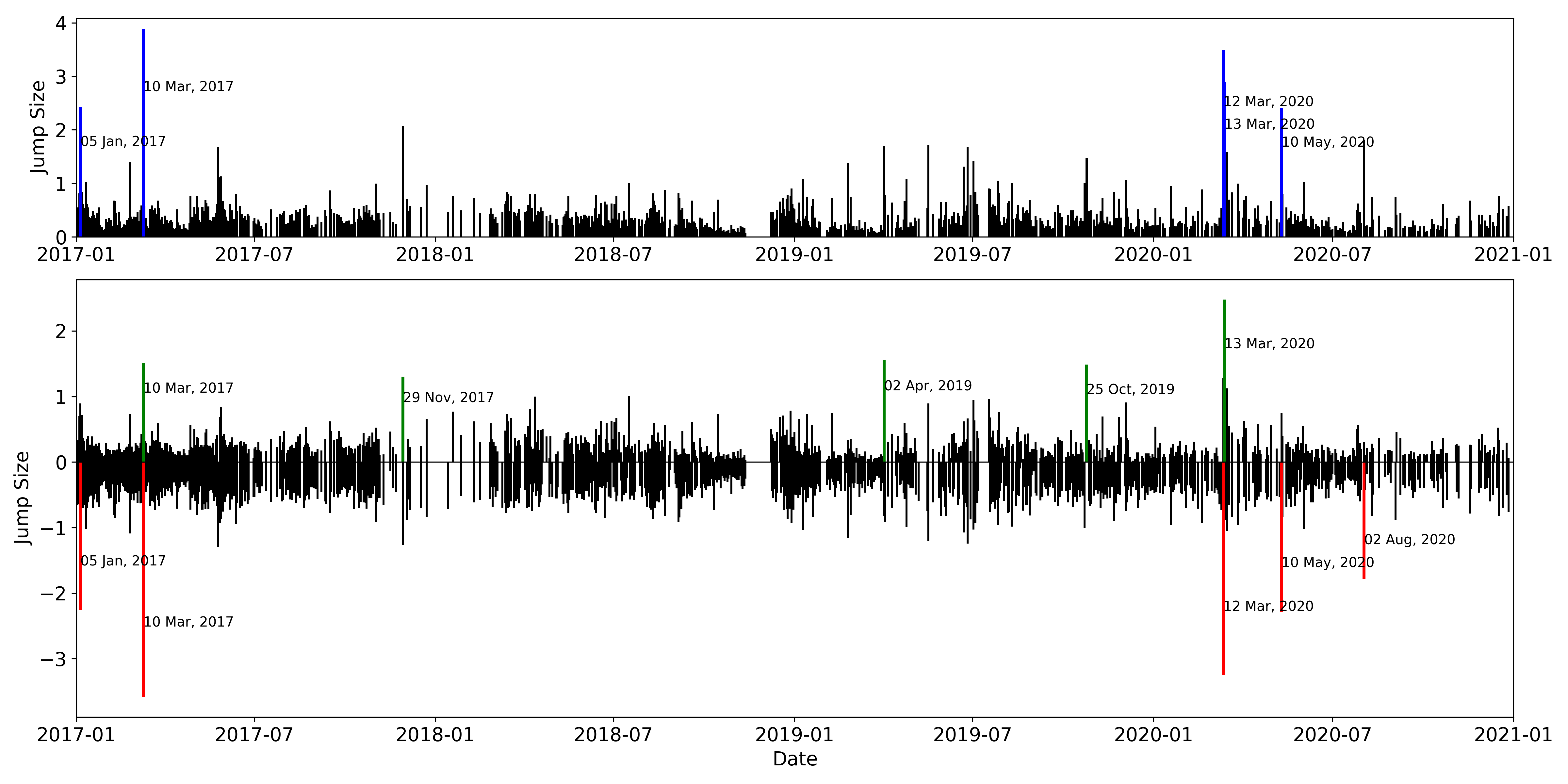}
      % \rule{\linewidth}{10em}
  \end{minipage}
  \caption{
      \textbf{Detected Price Jumps by Three Jump Estimators}
  \newline 
  \small
  The upper panel shows the jumps detected by the corrected jump estimator in the sample period, while the lower panel shows the jumps detected by signed jump estimators. The days detected with the largest 5 jumps are marked with specific dates.}
  \label{fig:jump_incidences}
\end{figure}

A sanity check on the detected jumps is conducted by corresponding large jumps with the market incidents. We demonstrate that the price jump of Bitcoin is not only the \enquote{sentimental} reaction to some of the opinion leaders like Elon Mask but majorly influenced by finance-related news. Jumps are usually considered as results from exogenous shocks, for example, the news shocks on the financial market. We try to link the news and reports to the detected jumps by exemplifying the top 5 jumps by each estimator. We choose five of the days detected with the largest price jumps marked in the first figure of the upper panel in Figure \ref{fig:jump_incidences}.

On 10th Mar. 2017, the Bitcoin ETF from Winklevoss twins was denied by the Securities Exchange Commission (SEC)\footnote{\enquote{Breaking: ETF Denied, Bitcoin Price Drops From \$1350 to \$980 Within Hours} from https://cointelegraph.com/}. As shown in the upper panel of Figure \ref{fig:jump_incidences}, that day is detected with the largest jump. Moreover, the lower panel of Figure \ref{fig:jump_incidences} indicates that there was a huge negative jump accompanied by a relatively large positive jump viewed as a price comeback.

From March to August 2020, Bitcoin Market has experienced an extremely volatile period, where we detect large positive jumps but, more significantly, many huge negative jumps. Such a volatile period is a mixed result of many macroeconomic incidents, including the global pandemic of COVID-19 causing a low expectation on the economy, the U.S Federal Reverse announcing an initiative of pumping \$1.5-trillion into financial market on 12th Mar. 2020\footnote{A report on the Bitcoin price plunge, see \enquote{Devastating Bitcoin Wipeout Could See The Price Go ‘Sub-\$1,000’} on Forbes.}, and also the Bitcoin-specific incidents, for example, an outage on a crypto exchange named Coinbase on 10th May 2020 causing a 10\% price fall in few minutes.

One can correspond more of the jumps to the news by simply searching online. For instance, On 29th Nov. 2017, BTC reached \$10,000 where many analysts suspected that the market would be more volatile caused by the profit-taking investors\footnote{Source by: https://www.fintechfutures.com/}. A mysterious bull run appeared on 12th Apr. 2018 and the BTC price hit \$8,011 from \$6,780 in a short time\footnote{https://www.investopedia.com/news/why-did-bitcoin-jump-1k-april-12/}. The point is that the jump estimators capture the news incorporated into the Bitcoin price and motivate us to investigate how the jumps impact the volatility in the future, which we discuss in the next section.

\section{Realized Variance Forecasting}\label{sec:forecasting}

\subsection{In-Sample Forecasting Analysis}

\begin{table}[!htb]
  \small
  \setlength{\tabcolsep}{0pt}
  \begin{threeparttable}
    \caption{Full-Sample Fitting Regression Results of Unsigned Estimators}\label{tab:InSampleReg_RVtype_BTCG}
    \begin{tabular*}{\linewidth}{@{\extracolsep{\fill}}>{\itshape}
      l *{2}{S[table-format=2.3,table-number-alignment = center]} 
      l *{2}{S[table-format=2.3,table-number-alignment = center]}
      l *{2}{S[table-format=2.3,table-number-alignment = center]}
      }
        \toprule\toprule
        
    \multicolumn{9}{l}{\textit{Dependent variable:}$\log(RV_{t:t+h})$} \\ 
    {}                                    & \multicolumn{2}{c}{$h=1$} & {}       & \multicolumn{2}{c}{$h=7$} & {}       & \multicolumn{2}{c}{$h=30$} \\
    \cline{2-3} \cline{5-6} \cline{8-9}
    {}    & \multicolumn{1}{c}{(1)}      & \multicolumn{1}{c}{(2)}      & {} & \multicolumn{1}{c}{(3)}     & \multicolumn{1}{c}{(4)}      & {} & \multicolumn{1}{c}{(5)}     & \multicolumn{1}{c}{(6)} \\
    \multirow{2}{*}{$\log(RV_{t-1:t})$}   & 0.489    & 0.512    & {} & 0.304   & 0.330    & {} & 0.141   & 0.119 \\
    {}                                    & (15.785) & (16.327) & {} & (8.13)  & (7.800)  & {} & (5.710) & (4.403) \\[0.15cm]
    \multirow{2}{*}{$\log(RV_{t-7:t})$}   & 0.254    & 0.265    & {} & 0.177   & 0.127    & {} & 0.120   & 0.091 \\
    {}                                    & (5.640)  & (5.288)  & {} & (2.382) & (1.499)  & {} & (1.681) & (0.997) \\[0.15cm]
    \multirow{2}{*}{$\log(RV_{t-30:t})$}  & 0.147    & 0.135    & {} & 0.236   & 0.293    & {} & 0.216   & 0.328 \\
    {}                                    & (3.299)  & (2.868)  & {} & (2.656) & (3.340)  & {} & (1.447) & (2.593) \\[0.15cm]
    \multirow{2}{*}{$\log(J_{t-1:t}+1)$}  & {}       & -0.383   & {} & {}      & -0.413   & {} & {}      & -0.096 \\
    {}                                    & {}       & (-3.038) & {} & {}      & (-3.238) & {} & {}      & (-1.347) \\[0.15cm]
    \multirow{2}{*}{$\log(J_{t-7:t}+1)$}  & {}       & -0.355   & {} & {}      & 0.329    & {} & {}      & 0.069 \\
    {}                                    & {}       & (-1.158) & {} & {}      & (0.705)  & {} & {}      & (0.130) \\[0.15cm]
    \multirow{2}{*}{$\log(J_{t-30:t}+1)$} & {}       & -0.300   & {} & {}      & -1.661   & {} & {}      & -3.109 \\
    {}                                    & {}       & (-0.784) & {} & {}      & (-2.261) & {} & {}      & (-3.248) \\[0.15cm]
    MZ-$R^2$                              & 0.237    & 0.211    & {} & 0.229   & 0.245    & {} & 0.238   & 0.329 \\
    % MSE                                   & 2.742    & 2.795    & {} & 1.098   & 1.062    & {} & 0.553   & 0.469 \\
    % HRMSE                                 & 1.029    & 0.965    & {} & 1.000   & 0.961    & {} & 0.959   & 0.865 \\
    % QLIKE                                 & 0.544    & 0.536    & {} & 0.713   & 0.693    & {} & 0.775   & 0.738 \\

          \bottomrule\bottomrule
        
    \end{tabular*}
  \begin{tablenotes}[flushleft]
    \setlength\labelsep{0pt}
  \linespread{1}\small
  \item 
  The table reports the full-sample regression results from HAR and RVJ models. Panels from left to right are the forecasting $RV$ in three different horizons, $h=1,7,30$. Each of the panel reports the estimated parameters of the two models. All parameters are estimated by OLS using Newey-West covariance matrix estimator with 7, 14 and 60 lags for $h=1,7,30$, respectively. The $t$-value is in the parentheses. The last row reports the $R^2$ of the Mincer-Zarnowitz regression.
  \end{tablenotes}
  \end{threeparttable}
\end{table}

\begin{table}[!htb]
  \small
  \setlength{\tabcolsep}{0pt}
  \begin{threeparttable}
    \caption{Full-Sample Fitting Regression Results of Signed Estimators}\label{tab:InSampleReg_SRVtype_BTCG}
    \begin{tabular*}{\linewidth}{@{\extracolsep{\fill}}>{\itshape}
      l *{2}{S[table-format=2.3,table-number-alignment = center]} 
      l *{2}{S[table-format=2.3,table-number-alignment = center]}
      l *{2}{S[table-format=2.3,table-number-alignment = center]}
      }
        \toprule\toprule
        
    \multicolumn{9}{l}{\textit{Dependent variable:}$\log(RV_{t:t+h})$} \\ 
    {}                                    & \multicolumn{2}{c}{$h=1$} & {}       & \multicolumn{2}{c}{$h=7$} & {}       & \multicolumn{2}{c}{$h=30$} \\
    \cline{2-3} \cline{5-6} \cline{8-9}
    {}    & \multicolumn{1}{c}{(1)}      & \multicolumn{1}{c}{(2)}      & {} & \multicolumn{1}{c}{(3)}     & \multicolumn{1}{c}{(4)}      & {} & \multicolumn{1}{c}{(5)}     & \multicolumn{1}{c}{(6)} \\
    \multirow{2}{*}{$\log(RSV^{+}_{t-1:t})$}  & 0.259    & 0.243    & {} & 0.169    & 0.259    & {} & 0.090    & 0.158 \\
    {}                                        & (5.466)  & (2.937)  & {} & (3.772)  & (3.479)  & {} & (3.234)  & (3.417) \\[0.15cm]
    \multirow{2}{*}{$\log(RSV^{+}_{t-7:t})$}  & 0.120    & 0.166    & {} & 0.018    & 0.032    & {} & -0.064   & -0.008 \\
    {}                                        & (1.147)  & (1.108)  & {} & (0.104)  & (0.127)  & {} & (-0.531) & (-0.045) \\[0.15cm]
    \multirow{2}{*}{$\log(RSV^{+}_{t-30:t})$} & 0.263    & 0.216    & {} & 0.422    & 0.397    & {} & 0.487    & 0.425 \\
    {}                                        & (2.283)  & (1.396)  & {} & (1.576)  & (1.201)  & {} & (1.331)  & (0.961) \\[0.25cm]
    \multirow{2}{*}{$\log(RSV^{-}_{t-1:t})$}  & 0.245    & 0.275    & {} & 0.148    & 0.082    & {} & 0.060    & -0.030 \\
    {}                                        & (5.503)  & (3.475)  & {} & (3.134)  & (1.168)  & {} & (2.032)  & (-0.670) \\[0.15cm]
    \multirow{2}{*}{$\log(RSV^{-}_{t-7:t})$}  & 0.120    & 0.090    & {} & 0.144    & 0.078    & {} & 0.169    & 0.091 \\
    {}                                        & (1.304)  & (0.643)  & {} & (0.949)  & (0.321)  & {} & (1.311)  & (0.490) \\[0.15cm]
    \multirow{2}{*}{$\log(RSV^{-}_{t-30:t})$} & -0.117   & -0.086   & {} & -0.185   & -0.108   & {} & -0.265   & -0.108 \\
    {}                                        & (-1.048) & (-0.53)  & {} & (-0.692) & (-0.312) & {} & (-0.737) & (-0.232) \\[0.25cm]
    \multirow{2}{*}{$\log(J^{+}_{t-1:t}+1)$}  & {}       & -0.133   & {} & {}       & -0.615   & {} & {}       & -0.346 \\
    {}                                        & {}       & (-0.499) & {} & {}       & (-2.616) & {} & {}       & (-2.671) \\[0.15cm]
    \multirow{2}{*}{$\log(J^{+}_{t-7:t}+1)$}  & {}       & -0.562   & {} & {}       & 0.234    & {} & {}       & -0.150 \\
    {}                                        & {}       & (-0.833) & {} & {}       & (0.203)  & {} & {}       & (-0.131) \\[0.15cm]
    \multirow{2}{*}{$\log(J^{+}_{t-30:t}+1)$} & {}       & -0.568   & {} & {}       & -2.141   & {} & {}       & -3.878 \\
    {}                                        & {}       & (-0.597) & {} & {}       & (-1.182) & {} & {}       & (-1.351) \\[0.25cm]
    \multirow{2}{*}{$\log(J^{-}_{t-1:t}+1)$}  & {}       & -0.314   & {} & {}       & 0.007    & {} & {}       & 0.209 \\
    {}                                        & {}       & (-1.259) & {} & {}       & (0.037)  & {} & {}       & (1.770) \\[0.15cm]
    \multirow{2}{*}{$\log(J^{-}_{t-7:t}+1)$}  & {}       & 0.047    & {} & {}       & 0.430    & {} & {}       & 0.256 \\
    {}                                        & {}       & (0.074)  & {} & {}       & (0.391)  & {} & {}       & (0.278) \\[0.15cm]
    \multirow{2}{*}{$\log(J^{-}_{t-30:t}+1)$} & {}       & 0.276    & {} & {}       & -0.082   & {} & {}       & -0.251 \\
    {}                                        & {}       & (0.302)  & {} & {}       & (-0.049) & {} & {}       & (-0.088) \\[0.15cm]
    MZ-$R^2$                                  & 0.227    & 0.211    & {} & 0.230    & 0.249    & {} & 0.248    & 0.342 \\
    % MSE    & 2.763 & 2.791 & {} & 1.095 & 1.057 & {} & 0.545 & 0.463 \\
    % HRMSE  & 0.997 & 0.959 & {} & 0.985 & 0.950 & {} & 0.911 & 0.874 \\
    % QLIKE  & 0.539 & 0.531 & {} & 0.718 & 0.693 & {} & 0.777 & 0.739 \\
    
          \bottomrule\bottomrule
        
    \end{tabular*}
  \begin{tablenotes}[flushleft]
    \setlength\labelsep{0pt}
  \linespread{1}\small
  \item 
    The table reports the full-sample regression results from RSV and RSVSJ models. Panels from left to right are the forecasting $RV$ in three different horizons, $h=1,7,30$. Each panel reports the estimated parameters of the two models. All parameters are estimated by OLS using Newey-West covariance matrix estimator with 7, 14, and 60 lags for $h=1,7,30$, respectively. The $t$-value is in the parentheses. The last row reports the $R^2$ of the Mincer-Zarnowitz regression.
  \end{tablenotes}
  \end{threeparttable}
\end{table}

To analyze how each explanatory variable affects future $RV$, we fit each forecasting model with the full sample, i.e., from the start of 2017 until the end of 2020, named full-sample forecasting.

Table \ref{tab:InSampleReg_RVtype_BTCG} reports the regression results of HAR and RVJ models, where the in-sample forecasting results show consistently that the lagged $RV$ has a strong and persistent positive relationship with future realized variance $\log(RV_{t:t+h})$ in all of the three forecasting horizons $h=1,7,30$, while the daily jump component has significant negative impact on future daily and weekly aggregated realized variance $\log(RV_{t:t+h})$ when $h=1,7$.
For instance, Column (1) in Table \ref{tab:InSampleReg_RVtype_BTCG}, the coefficient between the 1-day lagged realized variance and the $RV$ in the next day is $0.489$ with $t$-statistics=$15.785$, and the intensity and significance decrease with weekly lagged $RV$ and monthly lagged $RV$.
And this positive effect of lagged $RV$ decays with longer forecasting horizon, for example, in RVJ model, the impact from daily aggregated $RV$ on 1-day ahead, 7-day ahead, and 30-day ahead $RV$ are $0.512$ ($t$-statistics=$16.327$), $0.330$ ($t$-statistics=$7.800$), and $0.119$ ($t$-statistics=$4.403$). Also, the 1-day lagged detected jumps induce significantly lower future realized variances across all of the three forecasting horizons. As shown in Column (2), (4) of Table \ref{tab:InSampleReg_RVtype_BTCG}, the coefficients between daily jump size and future realized variances of 1-day ahead and 7-day ahead are $-0.383$ ($t$-statistics=$-3.038$) and $-0.413$ ($t$-statistics=$-3.238$). This suggests that the higher jump shocks actually reduce future risk.

Table \ref{tab:InSampleReg_SRVtype_BTCG} contains the in-sample forecasting results of the RSV and RSVSJ model, which account for the positive and negative effects from past realized variances or jumps.
Both positive and negative daily aggregated realized signed variance have a positive impact on the future realized variance across the three forecasting horizons. As shown in the first and fourth rows of Columns (1)-(6), all the coefficients of daily aggregated realized signed variances are significantly positive at the 1\% level. For instance, $\log(RV^+_{t-1:t})$ has a coefficient of $0.259$ ($t$-statistics=$5.466$) and $\log(RV^-_{t-1:t})$ has a coefficient of $0.245$ ($t$-statistics=$5.503$) to 1-day ahead $\log(RV_{t:t+1})$. However, the impacts from the weekly and monthly aggregated realized signed variances decrease, implying that the early realized signed variances are less informative.

The in-sample forecasting result also shows the importance of a positive jump on forecasting future realized variance, especially in the case of longer horizons. Column (4) and (6) in Table \ref{tab:InSampleReg_SRVtype_BTCG} shows that the a higher positive jump significantly reduces the future realized variances in the horizons of $h=7,30$. The coefficient of $\log(J^{+}_{t-1:t}+1)$ is $-0.615$ ($t$-statistics=$-2.616$) on $\log(RV_{t:t+7})$ and $-0.346$ ($t$-statistics=$-2.671$) on $\log(RV_{t:t+30})$. While the relationships in most cases between negative jumps and future realized variances are insignificant. This finding slightly differs from the result in \cite{patton2015good}, in which negative jumps lead to significant higher future volatility.

\subsubsection{Adaptive in Realized Variance Forecast}

The full-sample forecasting assumes the realized variance time series to be stable, while it is almost never true in an emerging market like the Bitcoin market. When the market changes intensively, any model could be biased when the calibrating period has a different structure than the forecasting period. Hence the fitted model would underperform in the out-of-sample forecast. We implement the rolling window method to allow the parameters to change over time, and then more reasonable comparisons can be obtained.

The adaptive method mimics an investor who updates the forecasting model based on the most recent information. A simple case is assuming that such updates are based on a fixed amount of lagged information.
The window size $T$ of the adaptive HAR models employed here is 90-days, i.e., models are estimated by using past 90-days samples. Moreover, all the models are re-estimated every day. After the re-estimation of each day, the out-of-sample forecasts are performed in horizons $h=1, 7, 30$, spontaneously.

\begin{figure}[!htb]
  \centering
  \begin{minipage}{1\linewidth}
      \includegraphics[width=\textwidth]{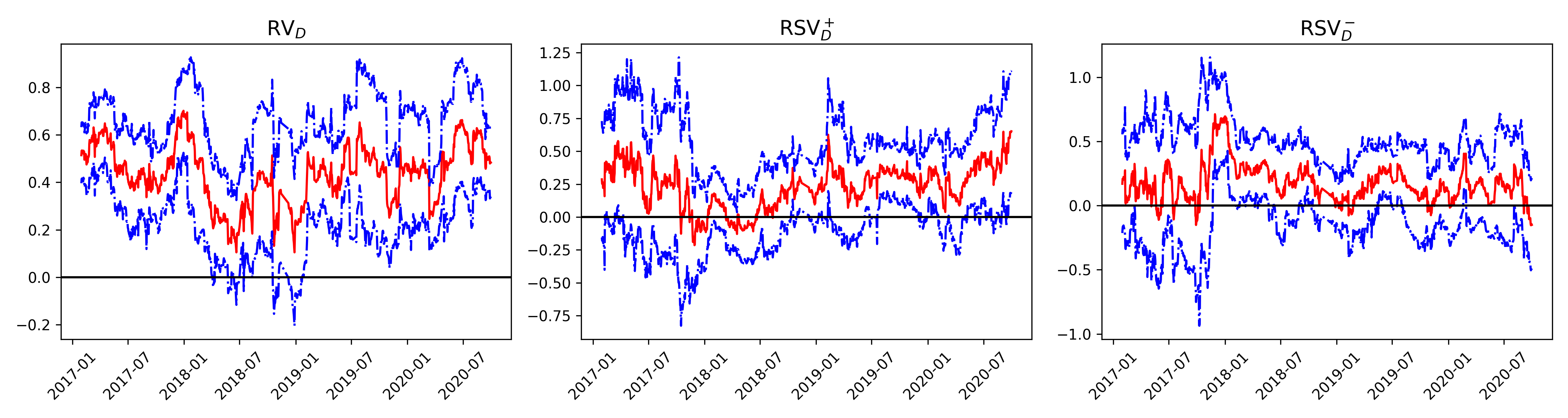}
      % \rule{\linewidth}{10em}
  \end{minipage}
  \caption{
      \textbf{Coefficients of Daily Aggregated Realized Variances}
  \newline 
  \small
  Time-varying parameters of the daily aggregated realized variance $\log(RV_{t-1:t})$ (left), positive realized variance $\log(RSV^+_{t-1:t})$ (middle), and negative realized variance $\log(RSV^-_{t-1:t})$ (right) in a 90-day rolling window forecast. Those three parameters are from HAR and RSV models with forecasting horizon $h=1$. Point estimated parameters in solid red line and the confidence interval with 95\% confidence level in blue dash line are reported in the period from Jan. 2017 until the end of 2020.
}
\label{fig:parameter_evolving_daily_RV}
\end{figure}

\begin{figure}[!htb]
  \begin{subfigure}[t]{1\textwidth}
    \includegraphics[width=\textwidth]{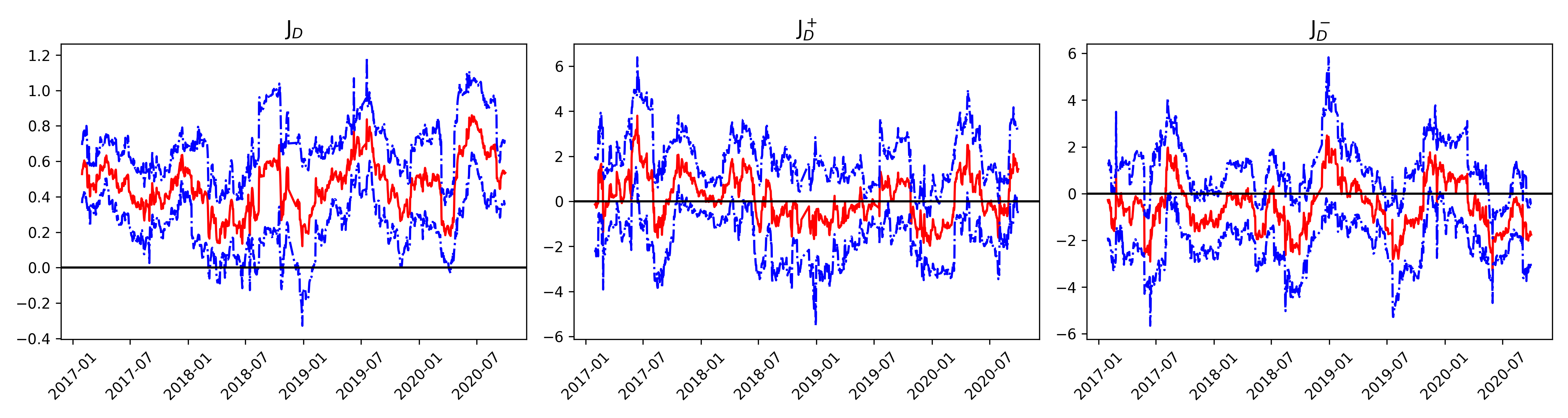}
    \caption{
      \textbf{$h=1$}
      % \newline 
      % \small
      }\label{fig:parameter_evolving_jump_h_1}
  \end{subfigure}
  \begin{subfigure}[t]{1\textwidth}
    \includegraphics[width=\textwidth]{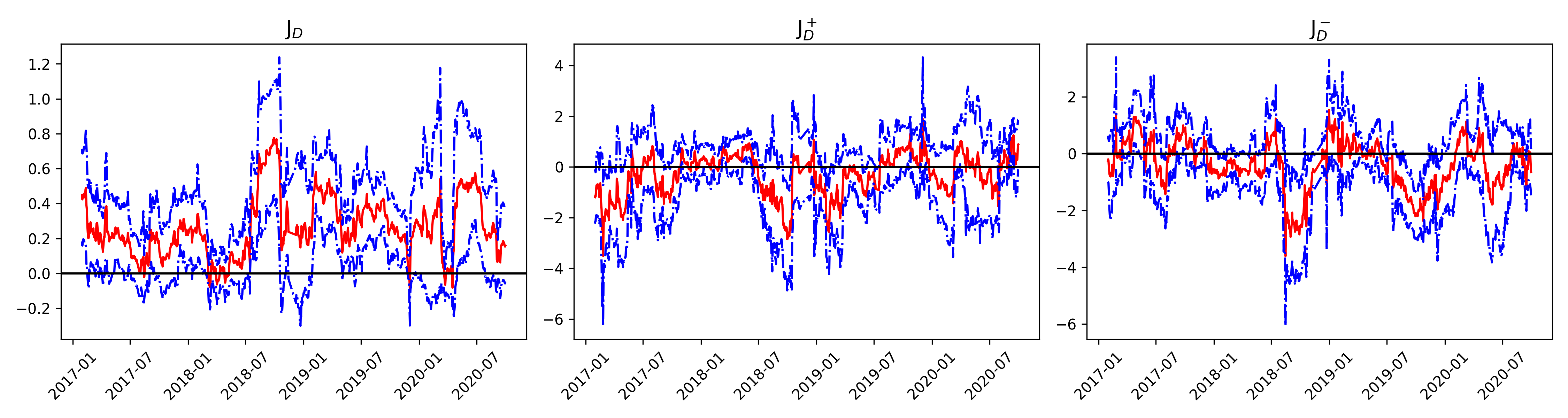}
    \caption{
      \textbf{$h=7$}
    %   \newline 
    %   \small
      }\label{fig:parameter_evolving_jump_h_7}
  \end{subfigure}
  \begin{subfigure}[t]{1\textwidth}
    \includegraphics[width=\textwidth]{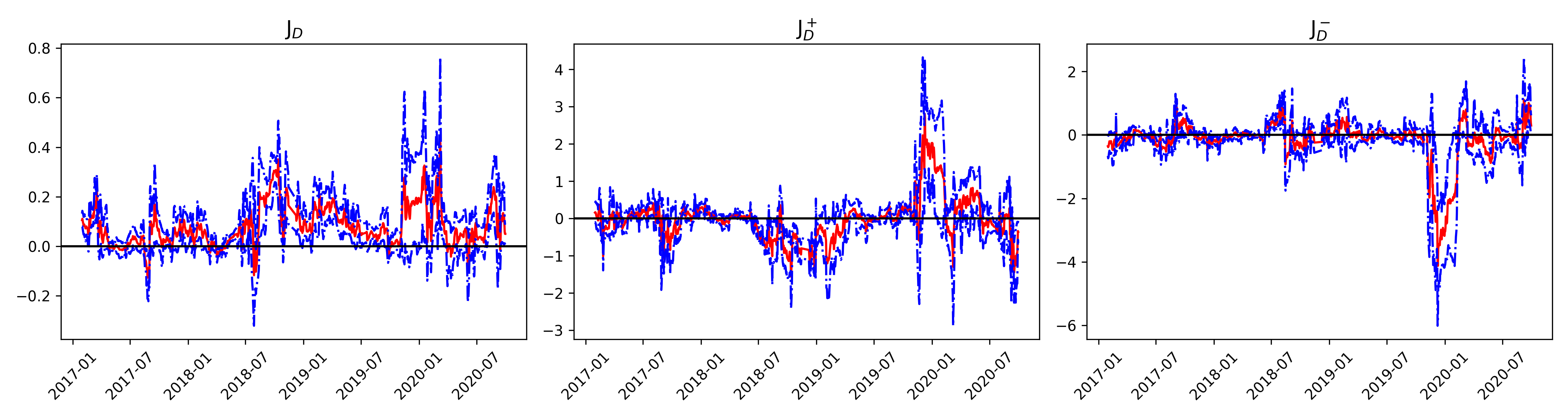}
    \caption{
      \textbf{$h=30$}
    %   \newline 
    %   \small
      }\label{fig:parameter_evolving_jump_h_30}
  \end{subfigure}
  \caption{
      % \textbf{Kernel Density Estimation and AutoCorrelation Function}
      % \newline 
      \small
      Each panel shows the time-varying parameters of the daily aggregated jump $\log(J_{t-1:t}+1)$ (left), positive jump $\log(J^{+}_{t-1:t}+1)$ (middle), and negative jump $\log(RSV^-_{t-1:t})$ (right) in a 90-day rolling window forecast. Point estimated parameters in solid red line and the confidence interval with 95\% confidence level in blue dash line are reported in the period from Jan. 2017 until the end of 2020.
      Panels (a) (b) and (c) show the results of forecasting horizon of $h=1,7,30$, respectively.
      }\label{fig:parameter_evolving_jump_h1_7_30}
\end{figure}

The parameters of the daily aggregated realized variances are evolving systematically, which justifies the adaptive forecasting method. Figure \ref{fig:parameter_evolving_daily_RV} shows the changing of parameters of the daily lagged realized variances $\log(RV_{t-1:t})$ and signed variances $\log(RSV^{+/-}_{t-1:t})$ in HAR and RSV model with forecasting horizon $h=1$. The solid red lines represent the point estimation of parameters, the blue lines are the confidence interval with a 95\% confidence level, and the black horizontal line indicates the zero value. The left figure confirms the significant positive impact of daily aggregated $RV$, which is persistent during the whole sample.
The other two figures in Figure \ref{fig:parameter_evolving_daily_RV} show that the upward and downward risk estimators play "complementary" roles to each other in forecasting over time. The upward risk coefficients went slowly down from 2017 until early 2018, and then it went for a trend of going up, compared with the pattern of the downward risk coefficients going opposite directions.

Figure \ref{fig:parameter_evolving_jump_h1_7_30} contains the coefficients changing of daily aggregated jump estimators. Panels from top to bottom are the different forecasting horizons $h=1,7,30$. In each panel, the coefficient of $\log(J_{t-1:t}+1)$ in the RVJ model is shown in the left figure, the coefficients of $\log(J^{+/-}_{t-1:t}+1)$ in the RSVSJ model are shown in the middle and right figure. Similar to the previous plot, the red dots are the point estimations on the coefficients in a 90-day rolling window, and the blue lines are the confidence interval with a 95\% confidence level.

The left figures of all three panels suggest that jumps will lead to a significantly higher future realized variances in different horizons $h=1,7,30$.  It is interesting to see that the positive and negative jumps significantly impact the 30-day ahead realized variance as shown in Panel (c) of Figure \ref{fig:parameter_evolving_jump_h1_7_30} during the start of the global pandemic in early 2020. In other words, the signed jumps are informative to forecasting long-term realized variance when the market is under extreme situation. In the shorter term realized variance forecast, the right two figures show that the parameters of signed jump estimators fluctuate below zero most of the time, indicating the negative but insignificant impact from signed jumps.

\subsection{Out-of-Sample Forecasts Evaluation}

In this subsection, we further discuss the out-of-sample forecasting results aiming to compare different models. All the out-of-sample forecasts are computed using 90-days rolling-window HAR regressions as described in the previous sections. Parameters are re-estimated on a daily basis. Here the "insanity filter" is applied in which we ensure that any forecast is no smaller (larger) than the minimum (maximum) realization of the past (\cite{patton2015good}, \cite{swanson1997forecasting} and \cite{bollerslev2018risk}).

\subsubsection{Forecasting Accuracy}

We employ multiple forecasting metrics to have a more robust comparison.
The out-of-sample performance evaluations are based on the squared form. In other words, forecasting results from the log-log forecasting models are transformed back to squared form for a fair comparison
\footnote{The squared form of HAR models is also common in previous literature and usually underperforms the logarithmic form. The squared form forecast unreported shows consistent results to the logarithmic form}.
We employ four metrics for the forecasting performance comparison.
The first one is $R^2$ from Mincer-Zarnowitz forecasting regression, named MZ-$R^2$.
The following three metrics named MSE, HRMSE, and QLIKE are computed from corresponding loss functions,

\begin{align}
  \operatorname{L^{MSE}}&=\left(RV_{t,t+h}-\widehat{RV}_{t,t+h}\right)^2\\\nonumber
  \operatorname{L^{HRMSE}}&=\left(\frac{RV_{t,t+h}-\widehat{RV}_{t,t+h}}{RV_{t,t+h}}\right)^2\\
  \operatorname{L^{QLIKE}}&=\log\widehat{RV}_{t,t+h}+\frac{{RV_{t,t+h}}}{\widehat{RV}_{t,t+h}},\nonumber
\end{align}

where $\widehat{RV}_{t,t+h}$ is the forecast average realized variance over period $[t,t+h]$, and $RV_{t,t+h}$ is the corresponding true value. The mean squared error (MSE) is the mean value of quadratic loss function $\operatorname{L^{MSE}}$ which measures the Euclidean distance between the ex-post realized variance and forecast result. The heteroscedasticity adjusted root mean squared error (HRMSE) is defined as the squared root mean of $\operatorname{L^{HRMSE}}$ (\cite{bollerslev1996periodic}) is a more robust metric to the scale changing of realized variance. The third metric, QLIKE, is the mean of a Gaussian quasi-likelihood loss function $\operatorname{L^{QLIKE}}$ (\cite{patton2011volatility}), which gives a consistent evaluation on different imperfect volatility proxies. Finally, the Diebold-Mariano test (\cite{diebold2002comparing}) is used to evaluate the significance of the difference between two forecasting results under the different metrics stated above.

\begin{table}[!htb]
  \renewrobustcmd{\bfseries}{\fontseries{b}\selectfont}
  \renewrobustcmd{\boldmath}{}
  \small
  \setlength{\tabcolsep}{0pt}
  \begin{threeparttable}
    \caption{Adaptive Log-Log HAR Models Out-of-Sample Forecasts Performance Evaluation}\label{tab:OutSampleForecasts_logform}
    \begin{tabular*}{\linewidth}{@{\extracolsep{\fill}}>{\itshape}
      l
      *{4}{S[table-format=1.3, table-number-alignment = center]}
      l
      *{4}{S[table-format=1.3, table-number-alignment = center]}
      }
      \toprule
      \toprule    
      \multicolumn{1}{c}{}
      & \multicolumn{4}{c}{MZ-$R^2$}
      & \multicolumn{1}{c}{}
      & \multicolumn{4}{c}{MSE}\\

      \cline{3-4}
      \cline{8-9}
    % \\
      \multicolumn{1}{c}{}
      &\multicolumn{1}{c}{HAR}
      &\multicolumn{1}{c}{RVJ}
      &\multicolumn{1}{c}{RSV}
      &\multicolumn{1}{c}{RSVSJ}
      &\multicolumn{1}{c}{}
      &\multicolumn{1}{c}{HAR}
      &\multicolumn{1}{c}{RVJ}
      &\multicolumn{1}{c}{RSV}
      &\multicolumn{1}{c}{RSVSJ}\\

      \cline{2-5}
      \cline{7-10}
      \\
    
      \textbf{h=1}  & 0.130 & 0.066 & \bfseries 0.177 & 0.111           & {} & 3.053    & 3.698    & \bfseries 2.864 & 3.207 \\
      \textbf{h=7}  & 0.223 & 0.224 & 0.261           & \bfseries 0.343 & {} & 1.154    & 1.186    & 1.106           & \bfseries 1.053 \\
      \textbf{h=30} & 0.400 & 0.462 & 0.533           & \bfseries 0.539 & {} & 0.462 & 0.412$*$ & 0.347$*$        & \bfseries 0.342$*$ \\
\\
      \multicolumn{1}{c}{}
      & \multicolumn{4}{c}{HRMSE}
      & \multicolumn{1}{c}{}
      & \multicolumn{4}{c}{QLIKE}
      \\
      \cline{3-4}
      \cline{8-9}
    % \\
      \multicolumn{1}{c}{}
      &\multicolumn{1}{c}{HAR}
      &\multicolumn{1}{c}{RVJ}
      &\multicolumn{1}{c}{RSV}
      &\multicolumn{1}{c}{RSVSJ}
      &\multicolumn{1}{c}{}
      &\multicolumn{1}{c}{HAR}
      &\multicolumn{1}{c}{RVJ}
      &\multicolumn{1}{c}{RSV}
      &\multicolumn{1}{c}{RSVSJ}    
      \\
      \cline{2-5}
      \cline{7-10}
      \\
      \textbf{h=1}  & 0.963 & 1.010$\dagger$ & \bfseries 0.947 & 1.176$\dagger$     & {} & \bfseries 0.689 & 0.745    & 0.700           & 0.916$\dagger$ \\
      \textbf{h=7}  & 0.898 & 1.047$\dagger$ & \bfseries 0.863 & 1.106              & {} & 0.800           & 0.832    & \bfseries 0.761 & 0.766 \\
      \textbf{h=30} & 0.766 & 0.674$*$       & 0.700           & \bfseries 0.668$*$ & {} & 0.730            & 0.690$*$ & 0.656$*$        & \bfseries 0.626$*$ \\

        \bottomrule
        \bottomrule
            
    \end{tabular*}
  \begin{tablenotes}[flushleft]
    \setlength\labelsep{0pt}
  \linespread{1}\small
  \item 
  The table reports the out-of-sample forecasting results by different accuracy metrics, Mincer-Zarnowitz $R^2$ (MZ-$R^2$), mean squared error (MSE), heteroscedasticity adjusted root mean squared error (HRMSE), and a Gaussian quasi-likelihood loss function (QLIKE). In each panel, the forecasting accuracy of four models of different forecasting horizons, $h=1,7,30$, is reported in different rows. The columns of each panel from left to right are HAR, RVJ, RSV, and RSVSJ model detailed in section \ref{subsec:har_model}. The model outperforms by absolute value in each category is marked in bold. The $\dagger$ marks when the HAR model outperforms the model significantly, while the $*$ denotes that the HAR model underperforms the model significantly, in which the significance is confirmed by the Diebold-Mariano test at 5\% significant level. All the forecasting models are in logarithmic form, and then the forecasting results are transformed back to squared form for the performance evaluation.
  
  \end{tablenotes}
  \end{threeparttable}
\end{table}

Contrary to \cite{andersen2007roughing}, we find that modelling separated jumps, or signed estimators do not necessarily improve the BTC realized variance forecasting accuracy all the time, but the forecasting horizon matters. The out-of-sample forecasting evaluation reported in Table \ref{tab:OutSampleForecasts_logform} shows clearly that forecasting performance depends on the forecasting horizon. This conclusion has a twofold meaning. First of all, it is obvious that each of the models performs better in the long forecasting horizon under most of the metrics. For instance, the MZ-$R^2$ is higher, and the MSE is much lower in $h=30$ compared with that in $h=1$. Secondly, it partially reveals the necessity of modeling jumps or decomposition into signed estimators. In the short forecasting horizon, $h=1$, for example, those models that do not include the separated jump components, HAR and RSV models, tend to outperform any other models. For example, MZ-$R^2$ is 0.130 in HAR, while it is 0.066 in RVJ, which accounts for the separated jump estimators. As shown in the first panel of Table \ref{tab:OutSampleForecasts_logform}, the Diebold-Mariano test shows the better accuracy of HAR at a 5\% significant level. However, in the long forecasting horizon case, $h=30$, modeling the jumps and signed estimators does improve the forecasting performance. As shown in the last row of each panel of Table \ref{tab:OutSampleForecasts_logform}, the models accounting jumps or signed estimators, RVJ, RSV, and RSVSJ, outperform the HAR model by all the metrics. The significance of outperformance of the RVJ and RSVSJ that models separated jumps components against the HAR model in the $h=30$ case is confirmed by the Diebold-Mariano test at 5\% significant level, as shown in the last row of Panel MSE, HRMSE, QLIKE of Table \ref{tab:OutSampleForecasts_logform}.

\subsubsection{Economic Value}\label{sec:utility}

From an economic point of view, it is necessary to address the importance of forecasting realized variance to Bitcoin investors. We employ the approach of \cite{bollerslev2018risk} to evaluate the utility that investors can obtain by forecasting realized variance. The advantages of this approach named Realized Utility relative to the framework of \cite{fleming2001economic} are twofold. Firstly, this RU framework evaluates utility without requiring forecasts on asset returns. Secondly, it mimics a trading strategy when an investor targets at a constant Sharp-ratio and adjusts his/her risky asset positions according to the $RV$ forecasts.

A first-order expansion on expected utility yields,

\begin{equation}
  \operatorname{E}\left[u(W_{t+1})\right] = \operatorname{E}(W_{t+1}) - \frac{1}{2}\gamma^{A}\operatorname{V}(W_{t+1}),
\end{equation}

where $\gamma^{A}$ is the absolute Pratt-Arrow risk aversion. The Realized Utility $RU_{t+1}^{(m)}$ in time period $(t:t+1]$ by model $m$ defined as utility per wealth with optimal weights $\operatorname{RU}^{(m)}_{t+1} = \operatorname{EU}(\omega_t^{(m)})/W_t$ is given by,

\begin{equation}
  \operatorname{RU}_{t+1}^{(m)} = \frac{SR^2}{\gamma}\left(\sqrt{\frac{RV_{t+1}}{\widehat{RV}_{t+1}^{(m)}}}-\frac{1}{2}\frac{RV_{t+1}}{\widehat{RV}_{t+1}^{(m)}}\right),
\end{equation}

where $RV_{t+1}$ and $\widehat{RV}_{t+1}$ are the ex-post and forecast realized variance of $t+1$.
Sharp ratio $SR=0.4$ and relative risk aversion $\gamma=2$ are given as constant.
If one has perfect forecast, i.e $\widehat{RV}_{t+1}^{(m)}=RV_{t+1}$, then $\operatorname{RU}_{t+1}^{(m)} = \frac{SR^2}{2\gamma} = 4\%$. See Appendix \ref{app_economic_value} for full details of realized utility.

As the optimal weight is given by \eqref{eq:optimal_omega} in Appendix \ref{app_economic_value}, for an investor given constant $SR/\gamma$, lower risk the investor expects for the next day, higher the proportion of wealth should be allocated to risky asset. We restrict the weight of Bitcoin as $\omega_t\in[0,1]$, i.e short-selling and leverage are not allowed. Therefore in the case of $\sqrt{\widehat{RV}_{t+1}^{(m)}}<SR/\gamma$, which means $\omega_t^{(m)}>1$, the realized utility $\operatorname{RU}^{(m)}_{t} = SR\cdot\sqrt{RV_{t+1}} - \frac{\gamma}{2}RV_{t+1}$, hence,

\begin{equation}\label{eq:ru}
  \operatorname{RU}^{(m)}_{t+1} =\begin{cases}
    \frac{SR^2}{\gamma}\left(\sqrt{\frac{RV_{t+1}}{\widehat{RV}_{t+1}^{(m)}}}-\frac{1}{2}\frac{RV_{t+1}}{\widehat{RV}_{t+1}^{(m)}}\right), & \text{$\frac{SR}{\gamma}\leq \sqrt{\widehat{RV}_{t+1}^{(m)}}$}\\
    SR\cdot\sqrt{RV_{t+1}} - \frac{\gamma}{2}RV_{t+1}, & \text{otherwise}
\end{cases}
\end{equation}

Clearly, the comparison almost solely depends on the forecasts $\widehat{RV}_{t+1}^{(m)}$ illustrated in \eqref{eq:ru}. Note that each of the forecasting models is in logarithmic form, and the realized utility is calculated after transforming the logarithmic forecast to squared form. The realized utility by squared form forecasting models unreported shows a consistent pattern.

The realized utility $RU^{(m)}$ for each model $m$ on forecasting horizon $h$ is computed by averaging the daily $RU^{(m)}_{t+1}$ over time period $(t:t+h)$,

\begin{equation}
  RU^{(m)}_h = \frac{1}{T} \sum_{t=1}^{T} RU^{(m)}_{t+h}
\end{equation}

The realized utility provides another metric to compare forecasts from different models. As explained above, better the forecast is, closer the $RU_{t+h}^{(m)}$ to $4\%$.

\begin{table}[!htb]
  \renewrobustcmd{\bfseries}{\fontseries{b}\selectfont}
  \renewrobustcmd{\boldmath}{}
  \small
  \setlength{\tabcolsep}{0pt}
  \begin{threeparttable}
    \caption{Economic Values Evaluation of Log-Log Models Out-of-Sample Forecasts (\%)}\label{tab:economic_evaluation_logform}
    \begin{tabular*}{\linewidth}{@{\extracolsep{\fill}}>{\itshape}
      l*{4}{S[table-format=1.3, table-number-alignment = center]}
      }
      \toprule
      \toprule    
      \multicolumn{1}{c}{}
      &\multicolumn{1}{c}{HAR}
      &\multicolumn{1}{c}{RVJ}
      &\multicolumn{1}{c}{RSV}
      &\multicolumn{1}{c}{RSVSJ}

      \\
      \cline{2-5}

      \textbf{h=1}  & 2.231 & 2.044 & \bfseries 2.249 & 1.811 \\
      \textbf{h=7}  & 2.580 & 2.550 & 2.717           & \bfseries 2.757 \\
      \textbf{h=30} & 3.342 & 3.450 & 3.551           & \bfseries 3.605 \\
    
        \bottomrule
        \bottomrule
            
    \end{tabular*}
  \begin{tablenotes}[flushleft]
    \setlength\labelsep{0pt}
  \linespread{1}\small
  \item 
  The table reports the economic value gained in terms of realized utility by forecasting realized variance with different models. The columns from left to right are realized utility from four different models, including HAR, RVJ, RSV, and RSVSJ model detailed in section \ref{subsec:har_model}. All the forecasting models are in logarithmic form, and the realized utility is calculated based on the forecasting results of the squared form. The rows are realized utility in three different forecasting horizons, $h=1,7,30$. The highest utility of each forecasting horizon is in bold. All the realized utilities are reported in percentage.
  \end{tablenotes}
  \end{threeparttable}
\end{table}

The economic value reported in Table \ref{tab:economic_evaluation_logform} confirms the conclusion from the out-of-sample forecasting. Firstly, an investor can gain higher utility regardless of the choice of the model if one forecasts in the long horizon risk. For instance, the RSVSJ model gives around 179 basis points more utility in the case of $h=30$ than that of $h=1$. 

Secondly, the forecasting horizon is still the key element in the decision of accounting for the separated jump components in the realized variance forecasting. On the one hand, in the short horizon forecasting, $h=1$, the models without separated jump components, HAR and RSV model, provide much more utility than the models that include jumps do. For example, HAR outperforms RVJ by up to around 19 basis points per day, suggesting that investors of Bitcoin who focus on short-horizon strategy should not account for the separated jump components in their forecasting models.  On the other hand, in the longer horizon forecasting case, $h=30$, accounting for either jumps or signed estimators, provide extra utility. For instance, the RVJ model outperforms the HAR model by 11 basis points utility per day, the RSV model outperforms HAR by around 21 basis points utility per day, and the RSVSJ model outperforms HAR by 26 basis points utility per day. This finding sheds light on realized variance forecasting models selection for BTC investors. BTC investors who target at a certain risk level should select the forecasting model based on their investment horizons.

   %!TEX root = ../RCVJ.tex

\section{Conclusion}\label{sec:conclusion}

As an emerging financial asset, Bitcoin has been booming in recent years and has become a major alternative asset for many investors worldwide. The risk characteristics of Bitcoin are the key to many of the future developments on Bitcoin.

This paper studies the risk of the Bitcoin market based on the high-frequency intraday data of a 3-year sample period from January 2017 to Doc. 2020. We find that the vanilla jump detection method suffers from the consecutive jumps, a commonly seen characteristic of Bitcoin but not well investigated in the previous empirical studies.

After correcting the bias caused by the consecutive jumps, the empirical evidence immediately shows that apart from the expected extreme high risk of Bitcoin, a large number of jumps entangle in the price process. We decompose positive and negative jumps by applying the realized semi-variance, which shows that the market surge and crash happened similarly in terms of average size and frequency. Further analysis shows that the price jumps link to many of the major economics-related occurrences, suggesting that jumps in Bitcoin reflect not only sentimental issues, e.g., opinion leaders' casual statements on Twitter, but also real-world incidents.

We employ 4 HAR-type models to study the forecasting properties of Bitcoin realized volatility. First of all, a full-sample forecasting result reveals that the 1-day lagged realized variance estimators and jump estimators impact the future realized variance significantly across the three forecasting horizons $h=1,7,30$. While among all the decomposed jump estimators, only the positive jump estimator negatively impacts the future 7-day and 30-day realized variance. Then, we allow the forecasting models to be adaptive with a 90-day rolling window. We find that the signed jumps can be a significant predictor of the future realize variance of the longer horizon, e.g., 30-day. Surprisingly, adding the separated jump estimators and the decomposed signed realized semi-variance explicitly into the forecasting models do not always improve the forecasting accuracy. It depends on the selection of the forecasting horizon. The out-of-sample forecasting results evaluated under several metrics reveal that an investor should model the jump estimator and the signed decomposed estimators when they are interested in forecasting the realized variance in a longer horizon, such as the 30-day realized variance. However, it does not provide extra forecasting accuracy or utility to model jumps for the investor focused on very short horizon realized variance like 1-day ahead realized variance.

  % Bibliography.
  \clearpage

  \bibliographystyle{jfe}
  \bibliography{ref}

\begin{thebibliography}{49}
\expandafter\ifx\csname natexlab\endcsname\relax\def\natexlab#1{#1}\fi

\bibitem[{Ait-sahalia et~al.(2005)Ait-sahalia, Mykland, and
  Zhang}]{ait2005often}
Ait-sahalia, Y., Mykland, P.~A., Zhang, L., 2005. How often to sample a
  continuous-time process in the presence of market microstructure noise. The
  Review of Financial Studies 18, 351--416.

\bibitem[{Andersen et~al.(2007)Andersen, Bollerslev, and
  Diebold}]{andersen2007roughing}
Andersen, T.~G., Bollerslev, T., Diebold, F.~X., 2007. Roughing it up:
  Including jump components in the measurement, modeling, and forecasting of
  return volatility. The Review of Economics and Statistics 89, 701--720.

\bibitem[{Andersen et~al.(2001{\natexlab{a}})Andersen, Bollerslev, Diebold, and
  Ebens}]{andersen2001distribution}
Andersen, T.~G., Bollerslev, T., Diebold, F.~X., Ebens, H., 2001{\natexlab{a}}.
  The distribution of realized stock return volatility. Journal of Financial
  Economics 61, 43--76.

\bibitem[{Andersen et~al.(2001{\natexlab{b}})Andersen, Bollerslev, Diebold, and
  Labys}]{andersen2001distributionexchangerate}
Andersen, T.~G., Bollerslev, T., Diebold, F.~X., Labys, P., 2001{\natexlab{b}}.
  The distribution of realized exchange rate volatility. Journal of the
  American Statistical Association 96, 42--55.

\bibitem[{Aït-Sahalia and Jacod(2012)}]{A_t_Sahalia_2012}
Aït-Sahalia, Y., Jacod, J., 2012. Analyzing the spectrum of asset returns:
  Jump and volatility components in high frequency data. Journal of Economic
  Literature 50, 1007--1050.

\bibitem[{Balcilar et~al.(2017)Balcilar, Bouri, Gupta, and
  Roubaud}]{balcilar2017can}
Balcilar, M., Bouri, E., Gupta, R., Roubaud, D., 2017. Can volume predict
  bitcoin returns and volatility? a quantiles-based approach. Economic
  Modelling 64, 74--81.

\bibitem[{Bandi and Russell(2008)}]{BANDI2008}
Bandi, F.~M., Russell, J.~R., 2008. Microstructure noise, realized variance,
  and optimal sampling. Review of Economic Studies 75, 339--369.

\bibitem[{Barndorff-Nielsen et~al.(2008{\natexlab{a}})Barndorff-Nielsen,
  Hansen, Lunde, and Shephard}]{barndorff2008designing}
Barndorff-Nielsen, O.~E., Hansen, P.~R., Lunde, A., Shephard, N.,
  2008{\natexlab{a}}. Designing realized kernels to measure the ex post
  variation of equity prices in the presence of noise. Econometrica 76,
  1481--1536.

\bibitem[{Barndorff-Nielsen et~al.(2008{\natexlab{b}})Barndorff-Nielsen,
  Kinnebrock, and Shephard}]{barndorff2008measuring}
Barndorff-Nielsen, O.~E., Kinnebrock, S., Shephard, N., 2008{\natexlab{b}}.
  Measuring downside risk-realised semivariance. CREATES Research Paper .

\bibitem[{Barndorff-Nielsen and
  Shephard(2002{\natexlab{a}})}]{barndorff2002econometric}
Barndorff-Nielsen, O.~E., Shephard, N., 2002{\natexlab{a}}. Econometric
  analysis of realized volatility and its use in estimating stochastic
  volatility models. Journal of the Royal Statistical Society: Series B
  (Statistical Methodology) 64, 253--280.

\bibitem[{Barndorff-Nielsen and
  Shephard(2002{\natexlab{b}})}]{barndorff2002estimating}
Barndorff-Nielsen, O.~E., Shephard, N., 2002{\natexlab{b}}. Estimating
  quadratic variation using realized variance. Journal of Applied Econometrics
  17, 457--477.

\bibitem[{Barndorff-Nielsen and
  Shephard(2004{\natexlab{a}})}]{barndorff2004econometric}
Barndorff-Nielsen, O.~E., Shephard, N., 2004{\natexlab{a}}. Econometric
  analysis of realized covariation: High frequency based covariance,
  regression, and correlation in financial economics. Econometrica 72,
  885--925.

\bibitem[{Barndorff-Nielsen and
  Shephard(2004{\natexlab{b}})}]{barndorff2004power}
Barndorff-Nielsen, O.~E., Shephard, N., 2004{\natexlab{b}}. Power and bipower
  variation with stochastic volatility and jumps. Journal of Financial
  Econometrics 2, 1--37.

\bibitem[{Barndorff-Nielsen and Shephard(2006)}]{barndorff2006econometrics}
Barndorff-Nielsen, O.~E., Shephard, N., 2006. Econometrics of testing for jumps
  in financial economics using bipower variation. Journal of Financial
  Econometrics 4, 1--30.

\bibitem[{Bollerslev and Ghysels(1996)}]{bollerslev1996periodic}
Bollerslev, T., Ghysels, E., 1996. Periodic autoregressive conditional
  heteroscedasticity. Journal of Business \& Economic Statistics 14, 139--151.

\bibitem[{Bollerslev et~al.(2018)Bollerslev, Hood, Huss, and
  Pedersen}]{bollerslev2018risk}
Bollerslev, T., Hood, B., Huss, J., Pedersen, L.~H., 2018. Risk everywhere:
  Modeling and managing volatility. The Review of Financial Studies 31,
  2729--2773.

\bibitem[{Bouri et~al.(2017)Bouri, Moln{\'{a}}r, Azzi, Roubaud, and
  Hagfors}]{Bouri2017}
Bouri, E., Moln{\'{a}}r, P., Azzi, G., Roubaud, D., Hagfors, L.~I., 2017. On
  the hedge and safe haven properties of bitcoin: Is it really more than a
  diversifier? Finance Research Letters 20, 192--198.

\bibitem[{Bukovina et~al.(2016)Bukovina, Marti{\v{c}}ek, and
  Others}]{bukovina2016sentiment}
Bukovina, J., Marti{\v{c}}ek, M., Others, 2016. Sentiment and bitcoin
  volatility. Tech. rep.

\bibitem[{Conrad et~al.(2018)Conrad, Custovic, and Ghysels}]{conrad2018long}
Conrad, C., Custovic, A., Ghysels, E., 2018. Long-and short-term cryptocurrency
  volatility components: A garch-midas analysis. Journal of Risk and Financial
  Management 11, 23.

\bibitem[{Corsi(2009)}]{corsi2009simple}
Corsi, F., 2009. A simple approximate long-memory model of realized volatility.
  Journal of Financial Econometrics 7, 174--196.

\bibitem[{Corsi et~al.(2010)Corsi, Pirino, and Reno}]{corsi2010threshold}
Corsi, F., Pirino, D., Reno, R., 2010. Threshold bipower variation and the
  impact of jumps on volatility forecasting. Journal of Econometrics 159,
  276--288.

\bibitem[{Diebold and Mariano(2002)}]{diebold2002comparing}
Diebold, F.~X., Mariano, R.~S., 2002. Comparing predictive accuracy. Journal of
  Business \& Economic Statistics 20, 134--144.

\bibitem[{Dyhrberg(2016)}]{Dyhrberg2016}
Dyhrberg, A.~H., 2016. Bitcoin, gold and the dollar {\textendash} a {GARCH}
  volatility analysis. Finance Research Letters 16, 85--92.

\bibitem[{Fan and Yao(2008)}]{fan2008nonlinear}
Fan, J., Yao, Q., 2008. Nonlinear Time Series: Nonparametric And Parametric
  Methods. Springer Science \& Business Media.

\bibitem[{Fleming et~al.(2001)Fleming, Kirby, and
  Ostdiek}]{fleming2001economic}
Fleming, J., Kirby, C., Ostdiek, B., 2001. The economic value of volatility
  timing. The Journal of Finance 56, 329--352.

\bibitem[{Garcia and Schweitzer(2015)}]{garcia2015social}
Garcia, D., Schweitzer, F., 2015. Social signals and algorithmic trading of
  bitcoin. Royal Society open science 2, 150288.

\bibitem[{Gerlach et~al.(2019)Gerlach, Demos, and
  Sornette}]{gerlach2019dissection}
Gerlach, J.-c., Demos, G., Sornette, D., 2019. Dissection of bitcoin’s
  multiscale bubble history from january 2012 to february 2018. Royal Society
  Open Science 6, 180643.

\bibitem[{Glaser et~al.(2014)Glaser, Zimmermann, Haferkorn, Weber, and
  Siering}]{glaser2014bitcoin}
Glaser, F., Zimmermann, K., Haferkorn, M., Weber, M.~C., Siering, M., 2014.
  Bitcoin-asset or currency? revealing users' hidden intentions. Revealing
  Users' Hidden Intentions (April 15, 2014). ECIS .

\bibitem[{Griffin and Shams(2018)}]{griffin2018bitcoin}
Griffin, J.~M., Shams, A., 2018. Is bitcoin really un-tethered? Avaiable at
  SSRN: https://ssrn.com/abstract=3195066 .

\bibitem[{Gronwald(2014)}]{gronwald2014economics}
Gronwald, M., 2014. The economics of bitcoins--market characteristics and price
  jumps. Available at SSRN: https://ssrn.com/abstract=2548999 .

\bibitem[{Hafner(2018)}]{10.1093/jjfinec/nby023}
Hafner, C.~M., 2018. Testing for bubbles in cryptocurrencies with time-varying
  volatility. Journal of Financial Econometrics .

\bibitem[{Hansen and Lunde(2006)}]{Hansen2006}
Hansen, P.~R., Lunde, A., 2006. Realized variance and market microstructure
  noise. Journal of Business {\&} Economic Statistics 24, 127--161.

\bibitem[{Hou et~al.(2018)Hou, Wang, Chen, and Härdle}]{svcj}
Hou, A.~J., Wang, W., Chen, C.~Y., Härdle, W.~K., 2018. Pricing cryptocurrency
  options: The case of bitcoin and crix. Available at SSRN:
  https://ssrn.com/abstract=3159130 .

\bibitem[{Huang and Tauchen(2005)}]{10.1093/jjfinec/nbi025}
Huang, X., Tauchen, G., 2005. {The Relative Contribution of Jumps to Total
  Price Variance}. Journal of Financial Econometrics 3, 456--499.

\bibitem[{Liu et~al.(2015)Liu, Patton, and Sheppard}]{liu2015does}
Liu, L.~Y., Patton, A.~J., Sheppard, K., 2015. Does anything beat 5-minute rv?
  a comparison of realized measures across multiple asset classes. Journal of
  Econometrics 187, 293--311.

\bibitem[{Mancini(2009)}]{mancini2009non}
Mancini, C., 2009. Non-parametric threshold estimation for models with
  stochastic diffusion coefficient and jumps. Scandinavian Journal of
  Statistics 36, 270--296.

\bibitem[{Nakamoto(2008)}]{nakamoto2008bitcoin}
Nakamoto, S., 2008. Bitcoin: A peer-to-peer electronic cash system .

\bibitem[{Nolte and Xu(2015)}]{nolte2015economic}
Nolte, I., Xu, Q., 2015. The economic value of volatility timing with realized
  jumps. Journal of Empirical Finance 34, 45--59.

\bibitem[{Patton(2011)}]{patton2011volatility}
Patton, A.~J., 2011. Volatility forecast comparison using imperfect volatility
  proxies. Journal of Econometrics 160, 246--256.

\bibitem[{Patton and Sheppard(2015)}]{patton2015good}
Patton, A.~J., Sheppard, K., 2015. Good volatility, bad volatility: Signed
  jumps and the persistence of volatility. Review of Economics and Statistics
  97, 683--697.

\bibitem[{Pichl and Kaizoji(2017)}]{bitcoin_volatility}
Pichl, L., Kaizoji, T., 2017. Volatility analysis of bitcoin price time series.
  Quantitative Finance and Economics 1, 474--485.

\bibitem[{Scaillet et~al.(2018)Scaillet, Treccani, and
  Trevisan}]{10.1093/jjfinec/nby013}
Scaillet, O., Treccani, A., Trevisan, C., 2018. {High-Frequency Jump Analysis
  of the Bitcoin Market*}. Journal of Financial Econometrics .

\bibitem[{Swanson and White(1997)}]{swanson1997forecasting}
Swanson, N.~R., White, H., 1997. Forecasting economic time series using
  flexible versus fixed specification and linear versus nonlinear econometric
  models. International Journal of Forecasting 13, 439--461.

\bibitem[{Traian~Pele et~al.(2019)Traian~Pele, Niels, H{\"a}rdle, Kolossiatis,
  and Yatracos}]{hardlephenotypic}
Traian~Pele, D., Niels, W., H{\"a}rdle, W.~K., Kolossiatis, M., Yatracos, Y.,
  2019. Phenotypic convergence of cryptocurrencies. IRTG 1792 Discussion Paper
  .

\bibitem[{Trimborn and H{\"a}rdle(2018)}]{trimborn2018crix}
Trimborn, S., H{\"a}rdle, W.~K., 2018. Crix an index for cryptocurrencies.
  Journal of Empirical Finance 49, 107--122.

\bibitem[{Urquhart and Zhang(2019)}]{urquhart2019bitcoin}
Urquhart, A., Zhang, H., 2019. Is bitcoin a hedge or safe haven for currencies?
  an intraday analysis. International Review of Financial Analysis 63, 49--57.

\bibitem[{Yermack(2015)}]{yermack2015bitcoin}
Yermack, D., 2015. Is bitcoin a real currency? an economic appraisal. In: {\em
  Handbook of digital currency\/}, Elsevier, pp. 31--43.

\bibitem[{Zhang(2006)}]{zhang2006efficient}
Zhang, L., 2006. Efficient estimation of stochastic volatility using noisy
  observations: A multi-scale approach. Bernoulli 12, 1019--1043.

\bibitem[{Zhang et~al.(2005)Zhang, Mykland, and Aït-sahalia}]{Zhang2005}
Zhang, L., Mykland, P.~A., Aït-sahalia, Y., 2005. A tale of two time scales.
  Journal of the American Statistical Association 100, 1394--1411.

\end{thebibliography}

  % Figures and tables, showing how to structure captions
  \clearpage

  %!TEX root = ../RCVJ.tex

\appendix

\section{}\label{app:risk_jump_data}  % Extra Empirical Results

\begin{table}[!htb]
  \small
  \setlength{\tabcolsep}{0pt}
  \begin{threeparttable}
    \caption{Summary Statistics of BTC Annualized Realized Variance Against Global Exchange Indices}\label{tab:des_indices_btc}
    \begin{tabular*}{\linewidth}{@{\extracolsep{\fill}}>{\itshape}
      l *{7}{S[table-format=2.2,table-number-alignment = center]}
      } 
      \toprule\toprule
      {}
      & \multicolumn{1}{c}{AEX$^{\dagger}$}
      & \multicolumn{1}{c}{DJI$^{\dagger}$}
      & \multicolumn{1}{c}{FTSE$^{\dagger}$}
      & \multicolumn{1}{c}{HSI$^{\dagger}$}
      & \multicolumn{1}{c}{SPX$^{\dagger}$}
      & \multicolumn{1}{c}{SSEC$^{\dagger}$}
      & \multicolumn{1}{c}{BTC-D}
      \\
      \midrule
      count & \num{4842} & \num{4704} & \num{4769} & \num{4645} & \num{4709} & \num{4508}  & \num{1385}\\
      mean  & 0.16       & 0.12       & 0.14       & 0.15       & 0.13       & 0.23        & 0.86   \\
      std   & 0.38       & 0.30       & 0.32       & 0.41       & 0.32       & 0.46        & 1.88   \\
      min   & 0.10\%     & 0.08\%     & 0.16\%     & 0.35\%     & 0.04\%     & 0.23\%      & 0.76\% \\
      25\%  & 0.02       & 0.02       & 0.02       & 0.03       & 0.02       & 0.03        & 0.16   \\
      50\%  & 0.05       & 0.04       & 0.05       & 0.06       & 0.05       & 0.09        & 0.37   \\
      75\%  & 0.14       & 0.11       & 0.13       & 0.14       & 0.12       & 0.23        & 0.85   \\
      max   & 7.04       & 5.55       & 7.74       & 16.46      & 7.18       & 7.71        & 36.93  \\
      \bottomrule\bottomrule
      \end{tabular*}
  \begin{tablenotes}[flushleft]
   \setlength\labelsep{0pt} 
  \linespread{1}\small
  \item $\dagger$: Selected global indices from developed markets and emerging markets. Trading hours in different global exchanges could be different which introduce bias of $RV$. We correct such bias by accounting the overnight price change (\cite{bollerslev2018risk}) to allow those $RV$ estimators to be comparable.
  \item Datasource from Realized Library, Oxford-Man Institute of Quantitative Finance.
\end{tablenotes}
  \end{threeparttable}
\end{table}

\begin{table}[!htb]
  \small
  \setlength{\tabcolsep}{0pt}
  \begin{threeparttable}
    \caption{Jump Estimators Summary Statistics For Synthetic Price Process}\label{tab:summary_btcd_jumps}
    \begin{tabular*}{\linewidth}{@{\extracolsep{\fill}}>{\itshape}
      l *{7}{S[table-format=1.3,table-number-alignment = center]}}
       
        \toprule
        \toprule
        \multicolumn{1}{c}{}
        & \multicolumn{1}{c}{$J_{u}(1-\alpha)$}
        & \multicolumn{1}{c}{$J(1-\alpha)$}
        & \multicolumn{1}{c}{$\log(J(1-\alpha)+1)$}
        & \multicolumn{1}{c}{$J^+$}
        & \multicolumn{1}{c}{$J^-$}
        & \multicolumn{1}{c}{$\log(J^+ +1)$}
        & \multicolumn{1}{c}{$\log(J^- +1)$}
        
        \\
        % \cline{1-2}
        \midrule
        {}  & \multicolumn{1}{c}{(1)}  & \multicolumn{1}{c}{(2)}  & \multicolumn{1}{c}{(3)}   
        & \multicolumn{1}{c}{(4)}   & \multicolumn{1}{c}{(5)}  & \multicolumn{1}{c}{(6)}     
        & \multicolumn{1}{c}{(7)}   \\[0.15cm] 

        prop      & 0.25    & 0.39     & {}        & {}          & {}          & {}           & {}   \\
        mean      & 0.31    & 0.37     & 0.30      & 0.25        & 0.26        & 0.21         & 0.22 \\
        std       & 0.17    & 0.26     & 0.17      & 0.21        & 0.23        & 0.15         & 0.16 \\
        min       & 0.09    & 0.10     & 0.09      & 0.03        & 0.03        & 0.03         & 0.03 \\
        5\%       & 0.11    & 0.12     & 0.11      & 0.06        & 0.07        & 0.06         & 0.07 \\
        50\%      & 0.26    & 0.30     & 0.26      & 0.17        & 0.19        & 0.16         & 0.17 \\
        95\%      & 0.63    & 0.80     & 0.59      & 0.70        & 0.65        & 0.53         & 0.50 \\
        max       & 0.81    & 2.26     & 1.18      & 1.24        & 2.09        & 0.81         & 1.13 \\
        skewness  & 0.86    & 2.48     & 1.43      & 2.09        & 2.96        & 1.58         & 1.73 \\
        kurtosis  & -0.08   & 10.73    & 3.17      & 5.06        & 15.08       & 2.50         & 4.72 \\
        acf(1)    & 0.19    & 0.19     & 0.23      & 0.10        & 0.11        & 0.13         & 0.15 \\
        acf(7)    & 0.15    & 0.14     & 0.17      & 0.10        & 0.11        & 0.13         & 0.13 \\

        \bottomrule
        \bottomrule
    \end{tabular*}
  \begin{tablenotes}[flushleft]
   \setlength\labelsep{0pt} 
  \linespread{1}\small
  \item Descriptive statistics of three jump estimation of a synthetic price process constructed by averaging price processes from multiple exchanges.
  Columns from left to right are the corrected thresholded jump estimator $J^{C}(1-\alpha)$ at $\alpha=99.99\%$ confidence level, the thresholded positive jump estimator $J^+$, and the thresholded negative jump estimator $J^-$.
  The threshold constant coefficient $c=3$.
  The first row reports the proportion of non-zero jumps.
  The following rows contain the sample mean, sample deviation, sample minimum, 50\% quantile, and sample maximum.
\end{tablenotes}
  \end{threeparttable}
\end{table}

\section{}  % Technical Details

\subsection{Convergence Properties of Realized Variance and Jump}\label{app:converge_prop_rvj}

Realized variance $RV_{t,t+1}$ is simply the cumulative squared logarithmic returns over time period $[t,t+1]$. By the theory of quadratic variation, the increment of Quadratic Variation $QV$ of $p(t)$ can be expressed as,

\begin{equation}
  \begin{split}
    QV_{t+1} &= \plim_{\Delta\to 0}\sum^{1/\Delta}_{j=1}r^2_{t+j\Delta}\\
             &= \int^{t+1}_{t} \sigma^2(s)ds + \sum_{t<s\leq t+1} \kappa^2(s)
  \end{split}
\end{equation}

The variation of $p(t)$ measured by $QV$ comes from two sources, one is driven by the c\`adl\`ag process and one is caused by the jump process. A series of literature discusses the convergence properties of $RV$. \cite{andersen2001distributionexchangerate}, \cite{barndorff2002econometric}, \cite{barndorff2002estimating} document the absence of jumps. Later, \cite{barndorff2004power}, \cite{barndorff2006econometrics}, \cite{andersen2007roughing} generalize to possible jumps. $RV$ converges in probability to $QV$ as $\Delta$ goes to 0,

\begin{equation}\label{eq:rv_decomp}
  RV_{t+1}(\Delta) \stackrel{\operatorname{p}}{\to} \underbrace{\int^{t+1}_{t} \sigma^2(s)ds}_{IV_{t+1}} + \underbrace{\sum_{t<s\leq t+1} \kappa^2(s)}_{J_{t+1}}
\end{equation}

Hence, $RV$ consists of two components: The continuous $IV$ component, and the Jump component $J_{u}$. The BiPower Variation $BPV$ measuring the continuous process allows separating the components,

\begin{equation}\label{eq:bpv}
  BPV_{t+1}(\Delta) = \mu_{1}^{-2}\sum_{j=2}^{1/\Delta} |r_{t+j\Delta}| \cdot |r_{t+(j-1)\Delta}|,
\end{equation}

where $\mu_1=\sqrt{2/\pi}$.

$BPV$ converges in probability to $IV$ in \eqref{eq:rv_decomp} as $\Delta$ goes to 0. Intuitively, $BPV$ is robust to an infrequent jump process as it is smoothed by accumulating the adjacent logarithmic returns. $J$ can, therefore, be isolated by taking the difference of $RV$ and $BPV$. And then the difference is truncated to guarantee that $J_{u}$ is non-negative.

\begin{align}\label{eq:bpv_j_conv_def}
  \begin{split}
      BPV_{t+1}(\Delta) &\stackrel{\operatorname{p}}{\to} \int^{t+1}_{t} \sigma^2(s)ds\\
      \vspace{-0.2cm}
      RV_{t+1}(\Delta) - BPV_{t+1}(\Delta) &\stackrel{\operatorname{p}}{\to} \sum_{t<s\leq t+1}\kappa^2(s)\\
      \vspace{-0.2cm}
      J_{t+1,u} &\stackrel{\operatorname{def}}{=} \max\left\{RV_{t+1}(\Delta) - BPV_{t+1}(\Delta),0\right\}
    \end{split}
\end{align}

\subsection{Thresholded Multi-Power Variance Estimators}\label{app:converge_prop_tmpv}

Similar to procedure for detecting the uncorrected jumps $J_{u}$, two special cases of $TMPV$ are used here. $TBPV$ estimates $\int_{t}^{t+1}\sigma^2(s)ds$ and $TTPV$ estimates $\int_{t}^{t+1}\sigma^4(s)ds$. Two estimators are defined as follows.
The general form of corrected $TMPV$ is defined as,

\begin{equation}
  TMPV_{t+1}(\Delta,\eta_1,\dots,\eta_m) \stackrel{\operatorname{def}}{=} \left(\prod_{k=1}^{m}\mu^{-1}_{\eta_k}\right) \cdot\Delta^{1-\frac{1}{2}(\eta_1+\dots+\eta_m)} \cdot\sum_{j=m}^{1/\Delta}\prod_{k=1}^{m}C_{\eta_k}\left(r_{t+(j-k+1)\Delta},\theta_{t+(j-k+1)\Delta}\right)
\end{equation}

We specify $m=2, \eta_1=\eta_2=1$ for $TBPV$, and $m=3, \eta_1=\eta_2=\eta_3=\frac{4}{3}$ for estimating $TTPV$, therefore we have,

\begin{alignat}{2}
    TBPV_{t+1}(\Delta)&=\mu_1^{-2}&&\cdot\sum_{j=2}^{1/\Delta}C_1(r_{t+j\Delta},\theta_{t+j\Delta})C_1(r_{t+(j-1)\Delta},\theta_{t+(j-1)\Delta})\\
    TTPV_{t+1}(\Delta)&=\mu_{\frac{4}{3}}^{-3}&&\cdot\Delta^{-1}\cdot\sum_{j=3}^{1/\Delta}C_{\frac{4}{3}}\left(r_{t+j\Delta},\theta_{t+j\Delta}\right)\\
                      &                        &&\cdot C_{\frac{4}{3}}\left(r_{t+(j-1)\Delta},\theta_{t+(j-1)\Delta}\right)\nonumber\\
                      &                        &&\cdot C_{\frac{4}{3}}\left(r_{t+(j-2)\Delta},\theta_{t+(j-2)\Delta}\right)\nonumber
\end{alignat}

Test for thresholded jumps $t\mbox{-}z$ is given by \eqref{eq:tj_test},
provided by \cite{corsi2010threshold} which is based on the ratio statistic from \cite{10.1093/jjfinec/nbi025}, detailed in \cite{andersen2007roughing} under continuous jump diffusion model. Where $\zeta=\frac{\pi^2}{4}+\pi-5$. Under a series assumptions, for the null hypothesis that no jump exists, $t\mbox{-}z$ converges to standard normal distribution as $\Delta$ goes to 0, i.e $t\mbox{-}z \stackrel{\mathcal{L}}{\rightarrow}N(0,1)$.

\begin{equation}\label{eq:tj_test}
  t\mbox{-}z_{t+1}=\frac{\left\{RV_{t+1}(\Delta)-TBPV_{t+1}(\Delta)\right\}RV_{t+1}^{-1}(\Delta)} {\sqrt{\Delta\cdot\zeta\cdot\max\left\{{1,\frac{TTPV_{t+1}(\Delta)}{\{TBPV_{t+1}(\Delta)\}^2}}\right\}}}
\end{equation}

\subsection{Conditional Expected Return}\label{app_conditional_return}

Essentially, instead of eliminating every of the points that has square-returns $r^2_{t+j\Delta}$ larger than certain threshold value $\theta_{t+j\Delta}$, the corrected $TMPV$ replaces the $\eta\mbox{-}th$ power logarithmic return $|r|^{\eta}_{t+j\Delta}$ with $r^e(\theta_{t+j\Delta},\eta)$ which is the expected value under assumption that $r_{t+j\Delta}\thicksim N(0,\sigma^2)$.
The conditional replacement logarithmic return $r^{C}_{\eta}(r_{t+j\Delta},\theta)$ can be written as:

\begin{equation}
  r^{C}_{\eta}(r_{t+j\Delta},\theta_{t+j\Delta}) = \left\{
            \begin{array}{lr}
            |r_{t+j\Delta}|^{\eta} & ,r^2_{t+j\Delta}\leq \theta\\
            r^e_{t+j\Delta}\left(\theta_{t+j\Delta}, \eta\right) & ,r^2_{t+j\Delta}>\theta
            \end{array}
          \right.
\end{equation}

The expected value of $\eta$-power returns conditioning on the square returns larger than threshold

\begin{align}
  \begin{split}
    r^e(\theta,\eta) &= \operatorname{E}\left\{|r|^{\eta} \bigg| r^2>\theta\right\}\\
                &=\frac{(2\sigma^2)^{\frac{\eta}{2}}}{2\sqrt{\pi}\Phi\left(-\frac{\sqrt{\theta}}{\sigma}\right)} \cdot\Gamma\left(\frac{\eta+1}{2},\frac{\theta}{2\sigma^2}\right)
  \end{split}
\end{align}

Given that the $\sigma^2$ is approximated by $\widehat{V_t}$, we have:

\begin{equation}\label{eq:replace_value}
  r^e(\theta_t,\eta) = \frac{1}{2\sqrt{\pi}\Phi\left(-c_{\theta}\right)} \cdot \left(\frac{2\theta_t}{c_{\theta}^2}\right)^{\frac{\eta}{2}} \cdot \Gamma\left(\frac{\eta+1}{2},\frac{c_{\theta}^2}{2}\right)
\end{equation}

Where $\Phi(x)$ is cdf of N(0,1) and $\Gamma(\alpha,x)=\int^{+\infty}_{x}s^{\alpha-1}e^{-s}ds$ is the upper incomplete gamma function.

\subsection{Local Variance Estimation}\label{app_local_V}

We employ the non-parametric local variance estimate \cite{fan2008nonlinear}

\begin{equation}\label{eq:local_var}
  \widehat{V}_{t}^{[n]}=\frac{\sum^{l}_{i=-l,i\neq{-1,0,1}}K(\frac{i}{l})\cdot r_{t+i}^2\cdot \mathbf{I}\{r_{t+i}^2\leq c_{\theta}^2\cdot \widehat{V}_{t+i}^{[n-1]}\}}{\sum^{l}_{i=-l,i\neq{-1,0,1}}K(\frac{i}{l})\cdot \mathbf{I}\{r_{t+i}^2\leq c_{\theta}^2\cdot \widehat{V}_{t+i}^{[n-1]}\}}, n=1,2,3...
\end{equation}

Where $K$ is a Gaussian kernel with bandwidth value $l=25$.
To avoid using future information and for computational simplicity, $\widehat{V}_t$ is estimated within each day. Thus, the first and last $l$-points of $\widehat{V}_t$ each day are smoothed by only partial Gaussian kernel. This recursive computation stops when the change from last step is smaller than a given criterion.

\subsection{Realized Utility}\label{app_economic_value}

A first order expansion on expected utility yields, for 1-day ahead,

\begin{equation}
  \operatorname{E}\left[u(W_{t+1})\right] = \operatorname{E}(W_{t+1}) - \frac{1}{2}\gamma^{A}\operatorname{V}(W_{t+1}),
  \label{eq:expect_utility}
\end{equation}

where $\gamma^{A}$ is the absolute Pratt-Arrow risk aversion.
The wealth function $W$ is explicitly given by \eqref{eq:wealth} for allocating $\omega_t$ proportion of whole wealth on the risky asset, and $r_{t+1}-r_f$ is the unknown excess return.

\begin{align}
  \begin{split}
    W_{t+1} &= W_t\{1 + (1-\omega_t) r_f + \omega_t r_{t+1}\}\\
            &= W_t\{1+r_f+ \omega_t(r_{t+1}-r_f)\}
    \label{eq:wealth}
  \end{split}
\end{align}

Assuming that the risk-free interest rate $r_f$ is constant, $W_t$ and $\omega_t$ are known, the expect value and variance of $W_{t+1}$ is:

\begin{align}
  \begin{split}
    \operatorname{E}(W_{t+1})&= W_t(1+r_f) + W_t\omega_t\operatorname(r_{t+1}-r_f)\\
    \operatorname{V}(W_{t+1})&= W_t^2\omega_t^2\cdot\operatorname{V}(r_{t+1}-r_f)
  \end{split}
  \label{eq:exp_var_w}
\end{align}

Given a target Sharp ratio $SR=\frac{\operatorname{E}(r_{t+1})}{\sqrt{\operatorname{V}(r_{t+1})}}$, \eqref{eq:exp_var_w} and \eqref{eq:expect_utility} give the following expression of expected utility $EU(\omega_t)$  $\operatorname{EU}(\omega_t)$ with replacing $\operatorname{V}(r_{t+1})$ by $RV_{t+1}$

\begin{align}
\operatorname{EU}(\omega_t) &= W_t\left[\omega_t\operatorname{E}(r_{t+1}) - \frac{\gamma}{2}\omega_t^2\operatorname{V}(r_{t+1})\right]\nonumber\\
    &= W_t\left[\omega_t\operatorname{E}(r_{t+1}) - \frac{\gamma}{2}\omega_t^2RV_{t+1}\right]\nonumber\\
    &= W_t\left[\omega_t\cdot SR\cdot\sqrt{RV_{t+1}} - \frac{\gamma}{2}\omega_t^2 RV_{t+1}\right]\label{eq:eu_final}
\end{align}

Here the $\gamma = \gamma^{A} W_{t}$ represents the relative risk aversion.
Based on the out-of-sample forecasts $\widehat{RV}_{t+1}^{(m)}$ from model $m$, one can derive the optimal weight $\omega_t^{(m)}$ targeting $SR/\gamma$.

\begin{equation}\label{eq:optimal_omega}
  \omega_t^{(m)} = \frac{SR/\gamma}{\sqrt{\widehat{RV}_{t+1}^{(m)}}}
\end{equation}

The Realized Utility $RU_{t+1}^{(m)}$ at time $t+1$ by model $m$ defined as utility per wealth with optimal weights $\operatorname{RU}^{(m)}_{t+1} = \operatorname{EU}(\omega_t^{(m)})/W_t$ can be obtained by \eqref{eq:optimal_omega} and \eqref{eq:eu_final}.

\begin{equation}
  \operatorname{RU}_{t+1}^{(m)} = \frac{SR^2}{\gamma}\left(\sqrt{\frac{RV_{t+1}}{\widehat{RV}_{t+1}^{(m)}}}-\frac{1}{2}\frac{RV_{t+1}}{\widehat{RV}_{t+1}^{(m)}}\right)\label{eq:eu_t}
\end{equation}

% \section{}

\clearpage

\end{document}